\documentclass[sigconf,screen]{acmart}

\AtBeginDocument{%
  \providecommand\BibTeX{{%
    \normalfont B\kern-0.5em{\scshape i\kern-0.25em b}\kern-0.8em\TeX}}}

\setcopyright{acmlicensed}
\copyrightyear{2018}
\acmYear{2018}
\acmDOI{XXXXXXX.XXXXXXX}

\acmConference[Conference acronym 'XX]{Make sure to enter the correct
  conference title from your rights confirmation email}{June 03--05,
  2018}{Woodstock, NY}
\acmISBN{978-1-4503-XXXX-X/18/06}

\usepackage{todonotes}
\usepackage{outlines}
\usepackage{multirow}
\usepackage[flushleft]{threeparttable}
\usepackage[inline]{trackchanges}

\usepackage{xspace}
\usepackage{booktabs}

\title[What Types of Questions Require Conversation to Answer?]{What Types of Questions Require Conversation to Answer? A Case Study of AskReddit Questions}

\definecolor{ao(english)}{rgb}{0.0, 0.5, 0.0}

\newcommand{\overview}[1]{}
\newcommand{\jack}[1]{}
 \newcommand{\review}[1]{}
\newcommand{\avon}[1]{}
\newcommand{\vivian}[1]{}
\newcommand{\td}[1]{}
\newcommand{\jc}[1]{}

\begin{document}

\title[Parental Collaboration and Closeness]{Parental Collaboration and Closeness: Envisioning with New Couple Parents}

\author{Ya-Fang Lin}
\email{yml5563@psu.edu}
\orcid{0000-0003-3689-5910}
\affiliation{%
  \institution{College of Information Sciences of Technology, Pennsylvania State University}
  \city{State College}
  \state{Pennsylvania}
  \country{USA}
}

\author{Xiaotian Li}
\email{xxl228@psu.edu}
\orcid{0009-0002-4802-4143}
\affiliation{%
  \institution{College of Information Sciences of Technology, Pennsylvania State University}
  \city{State College}
  \state{Pennsylvania}
  \country{USA}
}

\author{Wan-Hsuan Huang}
\affiliation{%
  \institution{Independent Researcher}
  \city{Warrensburg}
  \state{Missouri}
  \country{USA}
}
\email{037hsuang@gmail.com}
\orcid{0009-0001-1540-6615}

\author{Charan Pushpanathan Prabavathi}
\email{cjp6449@psu.edu}
\orcid{0009-0009-8889-235X}
\affiliation{%
  \institution{College of Information Sciences of Technology, Pennsylvania State University}
  \city{State College}
  \state{Pennsylvania}
  \country{USA}
}

\author{Jie Cai}
\authornote{Corresponding author}
\affiliation{%
  \institution{Department of Computer Science and Technology, Tsinghua University}
  \city{Beijing}
  \country{China}
  \postcode{100084}}
\email{jie-cai@mail.tsinghua.edu.cn}
\orcid{0000-0002-0582-555X}

\author{John M. Carroll}
\affiliation{%
  \institution{College of Information Sciences of Technology, Pennsylvania State University}
  \city{University Park}
  \state{Pennsylvania}
  \country{USA}
}
\email{jmc56@psu.edu}
\orcid{0000-0001-5189-337X}

\renewcommand{\shortauthors}{Lin et al.}

\begin{abstract}
  Couples often experience a decrease in closeness as they cope with the demands of parenthood. Existing technologies have supported parenting and parental collaboration. However, these technologies do not adequately support closeness in co-parenting. We use scenarios and design probes to brainstorm with 10 new parent couples to explore and envision possibilities for technologies to support closeness. We reported parents' current technology use for co-parenting and how participants considered and envisioned co-parenting technology for closeness, including information and task sharing, emotion awareness and disclosure, and fostering fun interaction. We discuss the potential technology has for fostering closeness in co-parenting by (1) fostering interdependence by supporting parental competence and (2) integrating positive emotions and experiences, such as validation and fun, in parenting. Based on our findings, we expand the design space of technology for closeness to include interdependence. We also expand the design space for co-parenting technology by integrating more positive emotions. \par

\color{red}{Preprint Accepted at ACM DIS 2025}
\end{abstract}




\maketitle

\section{Introduction}
Closeness refers to shared experiences, interdependence, and intimacy~\cite{berscheid1989relationship, rusbult2004interdependence, karandashev2019love}. Shared experiences involve spending time together~\cite{karandashev2019love, berscheid1989relationship}. Interdependence reflects the extent to which individuals rely on each other~\cite{rusbult2004interdependence}. Intimacy is conceptualized as self-disclosure and responsiveness~\cite{karandashev2019love}. 
A strong relationship can enable positive couple adaptation toward transition to parenthood~\cite{becotte2023positive}, while a lack of closeness can increase the risk of separation~\cite{berscheid1989relationship} and undermine well-being~\cite{deci2014autonomy, pietromonaco2017interpersonal}. 
Despite its importance,  couples sometimes find it difficult to maintain closeness. 

This paper focuses on fostering closeness during the transition of couples becoming new parents, a particularly difficult time for closeness maintenance. When new parents expand their roles to be not only a couple but also parents who collaborate for childcare, it is a critical, stressful, and unrelenting task. During this time, parenting leads new parents to exhaustion, lack of time for themselves, and increased disagreement~\cite{abreu2022couple}. They frequently struggle to balance the new and challenging collaborative responsibilities of taking care of their baby and maintaining their closeness. Described as ``Parenthood as crisis''~\cite{lemasters1957parenthood}, new parents experience a decrease in positive interchange and an increase in conflict~\cite{cowan1992partners, gottman2000decade}. When parents strive for the constant demands of their baby during sleepless nights or coordinate childcare responsibilities under huge pressure, they can sometimes ignore their partners' feelings, needs, or even efforts, not to mention having a movie night together. These experiences could undermine co-parents' closeness due to the lack of understanding and intimate interaction.
Furthermore, closeness and positive parental collaboration are bidirectionally associated~\cite{le2019cross}. When parents are closer, they collaborate in parenting better, and vice versa.
Given the importance and challenges of fostering closeness in co-parenting, designing collaboration technology that supports new parents' closeness becomes a critical need.

HCI scholars have studied various technological supports for parental collaboration, including coordination~\cite{nikkhah2022family, davidoff2010routine}, decision-making~\cite{kirchner2020just, grant2016intervention, yarosh2016best}, and providing support~\cite{song2020bodeum}. For example, \citet{davidoff2010routine} found most activities unfold as non-routine and suggested including routines in calendar systems for coordination. However, existing work mainly focuses on the logistics of parental collaboration, overlooking the need to support closeness.
Our research shifts to investigating the potential roles these technologies can play in enhancing closeness between co-parents in challenging co-parenting scenarios. We ask two research questions: 
\begin{itemize}
    \item \textbf{RQ1:} How do couples use technology in parental collaboration, including coordination, decision-making, and closeness?
    \item \textbf{RQ2:} How can technologies for co-parenting be designed, re-designed, and re-imagined to support closeness in co-parenting?
\end{itemize}

We held 10 initial sessions of co-design; each session with one unique pair of new parent couples. The participants shared their current technology use for co-parenting and envisioned how technology could foster closeness in co-parenting. From our qualitative analysis of the recordings, we find that participants used technology to share baby information and photos and coordinate childcare tasks. Yet, participants envisioned technology to foster closeness in co-parenting by motivating and facilitating information sharing and acquisition, enabling satisfactory task sharing, fostering emotional awareness and empathy, and enabling fun activities. Based on the findings, we propose design considerations and opportunities for fostering closeness in co-parenting, including enhancing interdependence and integrating positive emotions. In summary, this study contributes to both co-parenting and closeness in HCI, highlighting new couple parents' desires for parental collaboration technology to consider supporting closeness.

\section{Literature review}
\subsection{Closeness and technology supporting couples' closeness}

Social science studies have explored the meaning of closeness from various perspectives. Closeness has been constructed as behaviors, such as shared experiences~\cite{berscheid1989relationship, gottman1998psychology} and interdependence~\cite{rusbult2004interdependence, weingarten2022interdependence}; 
mental experiences, such as love~\cite{sternberg1986triangular}, identities~\cite{aron1997self}, and intimacy~\cite{reis2018intimacy}; and personal dispositions, such as childhood attachment history~\cite{hazan1994attachment}. This work uses closeness as a theoretically rich and sufficiently broad label. We refer to closeness as shared experiences, interdependence, and intimacy.

We conceptualize intimacy with Reis's interpersonal process model of intimacy~\cite{reis2018intimacy}, which explains a reciprocal process of how someone's self-disclosure (sharing personal thoughts, feelings, and experiences) and the partner's responsiveness (behaviors that express affection, empathy, and support) contribute to their intimacy. 
When people perceive that their partner's response to the disclosure is with understanding, validation, and care, they trust, like, and feel more intimate with the partner. This model also applies to married couples~\cite{laurenceau2005interpersonal}. 

We define interdependence as the extent to which individuals rely on each other~\cite{rusbult2004interdependence}. \citet{weingarten2022interdependence} interviewed two-profession couples and found the following words can convey quality interdependence: ``strength, sharing, mutual respect and regard, help, co-operation, dependence, reliance, activity, energy, taking over, picking up the slack, letting go, give and take, and willpower.'' 
For shared experiences, it describes behaviors such as how often partners engage in shared activities or show positive reciprocity in everyday conversations~\cite{berscheid1989relationship, gottman1998psychology}.

HCI scholars have investigated aspects of communication and interactions in close relationships. Solutions have focused on technology intervention to enhance intimacy and emotional connections, mostly among long-distance couples ~\cite{kowalski2013cubble, suzuki2017faceshare, eichhorn2008stroking, smith2018designing}. 
Some research investigated technology support closeness for collocated couples, focusing on understanding couples' intimacy~\cite{vetere2005mediating, branham2013designing}, couple collaboration~\cite{he2013couple}, ways couples explore, understand, and express desires~\cite{van2023like, ellegaard2021shaping}, and further facilitating spouses' presence and conversations~\cite{deuff2022together}. For example, \citet{he2013couple} found that intimate couples collaborate for experiences and affection rather than efficiency, like in the workplace.
\citet{lucier2018enhancing} designed an app for couples based on the relationship education model~\cite{futris2013national} to mediate self-disclosure and mutuality and further cultivate closeness between the couple.

While research has studied ways intimacy and communication contribute to a couple's closeness, it hasn't addressed scenarios where couples share critical, pressing, and unrelenting tasks, such as caring for a baby. Our research investigates ways technology can support co-parent collaboration in high-stakes situations during the early phases of parenthood and further cultivate closeness between the couple.

\subsection{Co-parenting in HCI} 
New parents need to form a solid dyad support system within the expanding family~\cite{cox1997families}. Parents need to be able to rely on each other to provide emotional, physical, and practical support. For example, when the father is sick and unable to pick up the baby from daycare, the mother makes a prompt decision to leave work early to pick up the baby. Co-parenting is used in a child-rearing context to indicate how parents collaborate, coordinate, support, and relate to each other in their parenting roles~\cite{mchale2004growing, feinberg2003internal}.
While co-parenting is commonly discussed among divorced parents, current family study scholars also emphasize the importance of co-parenting in resident families~\cite{campbell2023two}. The main components of co-parenting include supportive attitudes, closeness and solidarity, child-rearing agreements, the satisfaction of division of labor, and joint family management ~\cite{feinberg2002coparenting, feinberg2003internal, van2004coming}. 

Despite the joy of welcoming a newborn, many couples face challenges in their new co-parenting relationship, including a decline in marital satisfaction and mothers' frustration when expectations about the division of childcare responsibilities aren't met~\cite{twenge2003parenthood, christopher2015marital, khazan2008violated}. Both fathers and mothers perceived gendered norms of parenting roles~\cite{borgkvist2020unfortunately}. Research shows that fathers doubt the social appropriateness of their involvement in childcare~\cite{lukoff2017gender}. Mothers bore the dual responsibility of handling the majority of childcare and guiding their male partners on what care work to do, how to do it, and when~\cite{riggs2020s}. In addition, the lack of mutual understanding of parental needs and effective communication are also obstacles to positive co-parenting~\cite{lin2024ultimately}.

HCI scholars have studied ways technology could support parental collaboration by investigating parents' current practices in pregnancy care~\cite{lu2024unpacking}, sleep management~\cite{shin2022more}, medical or cultural child-rearing decision-making~\cite{kirchner2020just, yarosh2016best}, coordination for sickness and children's activities~\cite{nikkhah2022family, davidoff2010routine}, tension in technology use~\cite{derix2020probes}, and coordination after divorce~\cite{odom2010designing, dworkin2016coparenting}. For example, \citet{ko2022mobilizing} surveyed new mothers' practices and used technology to collaborate with stakeholders that provided childcare support. One design prototype explored ways to help parents reach a child development consensus~\cite{song2018bebecode}. Some research explored the potential for technology to facilitate empathy and respect parents' cultural experiences~\cite{yarosh2016best} and provide emotional support \cite{song2020bodeum}. However, existing research does not investigate parental collaboration through the lens of closeness. 

Couple relationships often become worse after the birth of a baby~\cite{shapiro2000baby}. Although research indicated that closeness and co-parenting are bidirectionally associated ~\cite{le2019cross}, few studies focused on ways technology could support both collaboration and closeness in co-parenting. 

Our study fills the gap by examining whether current technology supports or does not support co-parent interactions through the experience of new parents in resident families. New co-parents often experience an imbalanced division of labor~\cite{shin2022more, kirchner2020just, nikkhah2022family, jo2020understanding}. One study explored designs from mothers' perspective, which highlighted the need for greater spousal involvement in parenting ~\cite{jo2020understanding}. To address such challenges in co-parenting situations, we recruited both parents as a pair to collaborate and envision designs that meet their needs.

\section{Research context}
Our study focuses on the collaboration and closeness between new parents in Taiwanese families. In Taiwan, the beliefs of gender roles in parenting have shifted from traditional patriarchal beliefs to more progressive beliefs in recent years~\cite{ho2010parental}. For example, both mothers and fathers agree that fathers should not have more authority than mothers in decisions on children's education. However, mothers exhibit more progressive beliefs than fathers~\cite{ho2010parental}. For example, mothers believe they should work outside the home, whereas fathers remain neutral on this issue~\cite{ho2010parental}. 

Taiwanese fathers desire to be involved in family life, yet they predominantly see themselves as the "providers" for the family. While they actively engage in helping children with homework and participating in outdoor activities, their involvement in other aspects of parental duties is limited~\cite{hsiu2011parent}. Taiwanese mothers valued their partners' input in childcare, but first-time fathers did not have much to offer in solving childcare problems~\cite{huang2022maternal}.

The government established policies and provided resources that support equal gender responsibilities in new families. In Taiwan, companies are legally bound to offer two-year paternity leave for new fathers\footnote{Parental leaves in Taiwan, https://www.bli.gov.tw/0022925.html}. 
The Ministry of Health and Welfare in Taiwan introduced the Father's Parental Booklet in 2022 to support fathers in becoming parents\footnote{Father's parental booklet from the Ministry of Health and Welfare in Taiwan, https://www.hpa.gov.tw/Pages/Detail.aspx?nodeid=4576\&pid=15968}. 
However, over 80\% of Taiwanese fathers do not take parental leaves due to financial or career concerns, whereas mothers utilize maternity leaves more frequently~\cite{childwelfarefoundation2020father, ho2019father}. These gendered differences in taking parental leave for a newborn child are common in East Asian cultures~\cite{ho2019father}. The U.S. Census Bureau's data show that two-thirds of first-time fathers take time off after childbirth~\cite{USCensus2021}.

Many issues could affect a couple's relationship. Raising children is the most common issue reported by Taiwanese couples~\cite{pfeifer2013perceived}. This result differs from their Western counterparts; couples from the U.S. reported raising children as the least common issue affecting couples' relationships~\cite{su2015cross}. The closeness within couples is the most important factor that is positively related to marital satisfaction in Taiwanese couples, signaling a modernization of marital relationships from Traditional Chinese culture that devalues the importance of intimacy~\cite{shen2005factors}.

\section{Method}
To address our research questions, we conducted an initial session of co-design workshops with 10 new parent couples. Each session was conducted via Zoom using an interview approach in which one couple interacted with the first author, who was in the Zoom video call to facilitate the session. The design outcomes were the proposed designs that participants verbally shared with us and later identified in the data analysis process. We held 10 independent sessions, each with a unique couple participating. The duration of a session was approximately 1 to 1.5 hours. All sessions are conducted in Mandarin and recorded. The recordings were transcribed using speech recognition software and reviewed and analyzed by researchers.

\subsection{Recruitment and Participants}
In this section, we explain our study recruitment and participants, study procedure, design probes, qualitative analysis process, researcher positionality, and ethical considerations.
\subsubsection{Recruiting criteria and process}
The study was conducted in the middle of 2024.
Recruitment criteria included: (1) parents aged 18 years or older, (2) both parents must participate as a pair, and (3) the parents' first child must be younger than 4 years old.
We recruited parents whose first child was younger than 4 years old to ensure their first-time co-parenting experience was still fresh in their memory.
We recruited participants from social network groups in Taiwan, including Facebook groups for parents and Facebook groups where research opportunities are listed. We provided 800 NTD (around 25 USD) for each couple to thank them for their time.

\begin{table*}[]
\caption{Demographic information of participants}
\label{tab:participants}
\footnotesize
\begin{threeparttable}
\begin{tabular}{llllllll}
\hline
\textbf{ID} &
  \textbf{\begin{tabular}[c]{@{}l@{}}Year of \\ marriage\end{tabular}} &
  \textbf{Child age} &
  \textbf{\begin{tabular}[c]{@{}l@{}}Main caregiver \\ between the couple\end{tabular}} &
  \textbf{\begin{tabular}[c]{@{}l@{}}Other childcare support \\ (hours/weekday) \\ \end{tabular} } &
  \textbf{Age} &
  \textbf{Work hours/week} \\ \hline
P1M &
  \multirow{2}{*}{3y10m} &
  \multirow{2}{*}{2y1m; 2m} &
  \multirow{2}{*}{Even} &
  \multirow{2}{*}{9} &
  35 &
  not busy:20; busy: 60 \\
P1D &
   &
   &
   &
   &
  34 &
  40 \\ \hline
P2M &
  \multirow{2}{*}{4y} &
  \multirow{2}{*}{3y3m} &
  \multirow{2}{*}{Even} &
  \multirow{2}{*}{9} &
  33 &
  40-45 \\
P2D &
   &
   &
   &
   &
  33 &
  70 \\ \hline
P3M &
  \multirow{2}{*}{3y} &
  \multirow{2}{*}{9m} &
  \multirow{2}{*}{Mother} &
  \multirow{2}{*}{8} &
  36 &
  0 \\
P3D &
   &
   &
   &
   &
  35 &
  60 \\ \hline
P4M &
  \multirow{2}{*}{5y} &
  \multirow{2}{*}{2y6m} &
  \multirow{2}{*}{Mother} &
  \multirow{2}{*}{8} &
  34 &
  40 \\
P4D &
   &
   &
   &
   &
  37 &
  60 \\ \hline
P5M &
  \multirow{2}{*}{7y} &
  \multirow{2}{*}{3y4m; 1y10m} &
  \multirow{2}{*}{Even} &
  \multirow{2}{*}{11} &
  36 &
  45 \\
P5D &
   &
   &
   &
   &
  35 &
  40 \\ \hline
P6M &
  \multirow{2}{*}{10y} &
  \multirow{2}{*}{10m} &
  \multirow{2}{*}{Mother} &
  \multirow{2}{*}{10} &
  38 &
  40 \\
P6D &
   &
   &
   &
   &
  49 &
  40 \\ \hline
P7M &
  \multirow{2}{*}{5y} &
  \multirow{2}{*}{1y11m} &
  \multirow{2}{*}{Mother} &
  \multirow{2}{*}{8} &
  36 &
  30 \\
P7D &
   &
   &
   &
   &
  37 &
  40 \\ \hline
P8M &
  \multirow{2}{*}{2y6m} &
  \multirow{2}{*}{10 days} &
  \multirow{2}{*}{Even} &
  \multirow{2}{*}{0} &
  30 &
  40 \\
P8D &
   &
   &
   &
   &
  34 &
  40 \\ \hline
P9M &
  \multirow{2}{*}{4y} &
  \multirow{2}{*}{2y5m} &
  \multirow{2}{*}{Even} &
  \multirow{2}{*}{9} &
  30 &
  40 \\
P9D &
   &
   &
   &
   &
  32 &
  40 \\ \hline
P10M &
  \multirow{2}{*}{9y} &
  \multirow{2}{*}{10m} &
  \multirow{2}{*}{Mother} &
  \multirow{2}{*}{0} &
  38 &
  0 \\
P10D &
   &
   &
   &
   &
  40 &
  40 \\ \hline
\end{tabular}
\begin{tablenotes}
   \item Participant ID: M stands for mom, D stands for dad. e.g. P1M is P1, mom.
   \item Other childcare support includes grandparents and daycare.
   \item y: year; m: month; d: day

  \end{tablenotes}
  \end{threeparttable}
\end{table*}

\subsubsection{Participants}
All participants were cohabiting Taiwanese couples. The participants' interest in parental collaboration ranged from somewhat interested to very interested, based on a scale including: not interested, somewhat uninterested, neutral, somewhat interested, and very interested. Children's ages ranged from 10 days to 3 years and 4 months. Participants' ages ranged from 32 to 49 years. All participants had a Bachelor's degree or a Graduate degree. Two participants were homemakers (P3M, P10M). The rest were knowledge workers (e.g., legal professionals, freelancers, researchers, engineers, sales, marine surveyors, designers, and therapists), with weekly working hours ranging from 30 to 70 hours.
In terms of caregiving, five couples reported sharing the main caregiver role equally, while the other five indicated that the mother was the primary caregiver. P8 and P10 had no additional childcare support. P5 relied on the baby's grandparents for 11 hours of care on weekdays, and the remaining couples used daycare for 8 to 10 hours on weekdays. Please see Table \ref{tab:participants} for detailed information.

\subsection{Study Procedures and Design Probes}
Co-design is a human-centered design (HCD) approach that enables diverse voices to collaboratively envision and imagine in the design process~\cite{steen2013co}. HCI researchers frequently use co-design methods to develop digital technologies centered on end-users' perspectives~\cite{jo2020understanding, wardle2018exploring}. With the research goal to understand new parent couples' desires and needs of co-parenting technology for closeness, co-design is a proper research method as users are experts in their experiences~\cite{sanders2008co}. 

We prepared design probes to enable participants to envision designs from their lived experiences. First, we prepared problematic co-parenting situations for participants to easily identify their past co-parenting scenarios that they wanted to design. Using participants' scenarios, we aimed to enable participants to easily engage in design via scenarios that were vivid to them~\cite{carroll2003scenario}. Similarly, we prepared existing app functions for co-parents and couples to provide a starting point for participants to brainstorm solutions. In addition, we arranged stakeholders (i.e., father and mother in our study) as a unit to collaboratively ideate design solutions that fulfill the desires of both~\cite{lindquist2007co}. Below, we introduce the study procedure and design probes.

\subsubsection{Study Procedure}
Each session included two activities. The first activity was a semi-structured interview to explore the couple's experiences with existing technologies for co-parenting. Participants were asked about their current technology use for co-parenting, considerations for selecting these tools, and moments in co-parenting that foster or hinder closeness.
The second activity was an envisioning activity, where the couple collaborated to envision and design desired technology functionalities to enhance closeness in co-parenting.
The envisioning activity comprised three steps:
\begin{itemize}
    \item \textbf{Step 1: Select co-parenting scenarios.} 
    We presented a list of common problematic co-parenting situations and asked participants to select one they had experienced or could relate to. Participants could also propose additional scenarios that are not listed.

    \item \textbf{Step 2: Introduce existing app functions.} 
    We introduced existing app functions for co-parenting and couples to inspire participants, explaining that the showcased design features served as references to guide brainstorming. The screenshots of app functions are shown on the screen as a visual aid, and the researcher explained the use of each function. All screenshots are shown in English and verbally explained in Mandarin.

    \item \textbf{Step 3: Brainstorm and discuss technological solutions.}
    We then invited participants to share one to two examples of their experiences with the selected co-parenting situation as problem scenarios~\cite{carroll2003scenario} for brainstorming. We used various prompts to facilitate a discussion to envision helpful functionalities to maintain closeness during the selected scenarios. An example of prompt questions was ``How could you use or modify the design features presented or other technologies that you have used or known before to support you to feel closer in this scenario? Why?''. The participants discussed the functions that stood out to them and envisioned their designs based on them. Finally, we explored trade-offs between efficiency, fairness, and closeness by asking participants to navigate the function's influence on the three considerations.
\end{itemize}
At the end of each session, we collected demographic information and expressed our gratitude to the participants.

\subsubsection{Design probes}
\paragraph{Problematic co-parenting situations}
We draw problematic co-parenting scenarios from previous literature that investigated new parents' co-parenting practices and challenges (e.g. \cite{setiawan2022understanding}, \cite{kwon2013mothers}, \cite{christopher2015marital}, and \cite{sheedy2019coparenting}). The situations we provide were: (1) the child treats two parents differently, (2) childcare task rearrangement, (3) empathy about each other's difficulties, (4) overwhelming childrearing tasks, (5) a lack of perceived validation of parenting efforts, (6) emotion management; (7) initiation of a difficult conversation, (8) information imbalance, and (9) a lack couple time. The participants could use the provided problematic co-parenting situations as a starting point to create or extend scenarios based on their own experiences (e.g., distant moments they had mentioned) or those they found most relatable.

\paragraph{Existing app functions for co-parents and couples}
We used app functions related to parental collaboration and couple communication as technology probes to inspire parents to think about new technologies. We chose phone apps because parents could be more familiar with apps than other technology, as they commonly used apps for collaboration and communication~\cite{ko2022mobilizing}.
We collected functions from popular, highly-rated commercial co-parenting, couple, pregnancy, or baby apps as they are related to collaboration, connection, and support in parental collaboration and closeness. Such apps included BabySparks - Development App~\footnote{\url{https://apps.apple.com/us/app/babysparks-development-app/id794574199}}, WeParent - Co-Parenting App~\footnote{\url{https://apps.apple.com/us/app/weparent-co-parenting-app/id1441850251}}, TimeHut - Baby Album~\footnote{\url{https://apps.apple.com/us/app/timehut-baby-album/id565951606}}, Pregnancy \& Baby Tracker -WTE~\footnote{\url{https://apps.apple.com/us/app/pregnancy-baby-tracker-wte/id289560144}}, Gottman Card Decks~\footnote{\url{https://apps.apple.com/us/app/gottman-card-decks/id1292398843}}, and Love Nudge~\footnote{\url{https://apps.apple.com/us/app/love-nudge/id495326842}}. 

We categorized the functions of these apps into the following categories: shared calendar, shared contacts, shared to-do list, done list, reminder, finance, messaging and communication, baby activity tracking and information, parents' journal, and parental guidance and practices. The final visualization of the technology probes was a list of these co-parenting functions, accompanied by corresponding screenshots from the apps.

\subsection{Data analysis}
We imported all de-identified transcripts into Atlas.ti \footnote{\url{https://atlasti.com}}, a qualitative data analysis tool for four researchers to code collaboratively. We used inductive coding to identify concepts from the transcripts and grouped related concepts into broader themes. We also identified the participants' proposed designs, using codes such as ``D: design for shared responsibility/task'' and ``D: design to show mood/emotion''. At the beginning of the process, the first author reviewed all transcripts to generate a general data summary aligned with the research interests and shared it with the other three researchers to ensure a shared understanding of the research goals. Next, each of the four researchers independently coded one transcript containing abundant content. Following this initial coding, the researchers convened in a group meeting to discuss the codes, clarify definitions, and reach a consensus on the initial codebook. All codes, along with their definitions, were archived. For the remaining transcripts, three researchers independently read and coded two transcripts each, adding new codes as necessary. Once all transcripts were coded, the team exported 135 codes into a spreadsheet and iteratively organized the relevant codes into subcategories, categories, and high-level themes. Finally, all the quotes are translated from Mandarin to English.

\subsection{Researcher Positionality}
Our research team includes a new mother, a new father, a senior father, and childless unmarried individuals. All authors identify as East Asian, South Asian, or White. The academic background of the researchers includes sociology, psychology, counseling, education, and information science.
Researchers who are new parents provide us with insider perspectives, such as direct experiences using technology to co-parent and experiences about co-parenting and closeness. In contrast, non-parent researchers provide outsiders' viewpoints. 
We acknowledge our personal experiences and positionality as researchers, recognizing them as crucial points of reflection, given that \textit{``the subjective and objective components of knowledge are interconnected and interactive''}~\cite{banks1998lives}.
Thus, we carefully document every phase of our research process to provide sufficient details for readers to contextualize our findings and interpretations within a socio-cultural context.

\subsection{Ethical Considerations}
The study was reviewed and approved by our Institutional Review Board. We required oral informed consent from all participants to participate in
this study. We informed our participants that they could pause or withdraw from the study at any time without any penalties. The interviewer received training on navigating sensitive topics and addressing participants' emotions to minimize potential harm.

\section{Findings}
In our effort to explore how technology can facilitate closeness within couples, we begin by examining couples’ lived experiences: how parental collaboration affects relational closeness (Section \ref{How parental collaboration affects couple closeness}), and how technology currently plays a role in that collaboration (Section \ref{How do couples use technology in parental collaboration}). These sections contextualize the everyday realities and tensions that shape participants’ perspectives and thus serve as the foundation for understanding the technology designs they later envisioned. Building on this grounding, Section \ref{How to better design co-parenting technology for closeness} presents participants’ design ideations for how technology might support closeness in parental collaboration.

\subsection{How parental collaboration affects couple closeness?}\label{How parental collaboration affects couple closeness}

In this section, we describe how parents perceived their relational closeness to be shaped by the dynamics of parental collaboration. 

\subsubsection{Parental Collaboration Increases Couple's Distance.}\label{Distant Moments in Parental Collaboration}
Our participants shared with us various distant moments. Parents reported feeling distant when they experienced conflicts over the division of labor, differences in child-rearing values between partners or with grandparents, and a lack of time to connect with their partners amid the demands of their new roles as parents.

As the child became the focal point of the family, the time and activities that couples previously enjoyed together were restructured into family-oriented tasks, leaving little to no time for partners to connect with each other: 
\begin{quote}
\textit{``We can't watch TV together or do things like that anymore... 
Now, one person has to put the baby to sleep while the other watches TV.''} -P9D
\end{quote}
In this case, the P9 couple could not find a new way to pay attention to each other and cultivate their relationship due to the high demand of a newborn baby.

The most salient reason for feeling distant, as reported by our participants, was that mothers assumed the primary caregiving role, while fathers took a secondary role. This imbalance in childcare responsibilities frequently reduced closeness within couples. As P6M shared:
``\textit{
    That feeling of being alone in the middle of the night, trying to soothe the baby for one or two hours with no one there to help—it really damages the relationship between husband and wife.}''
Some mothers shared that they took on more childcare tasks because they were more experienced, and it was easier to complete the tasks themselves instead of teaching or waiting for their husbands to do them. 
However, the tendency to immediately address such tasks often led to a disproportionate burden of domestic responsibilities. 

Fathers in our study often expressed a sense of guilt and incompetence, or reported feeling pushed out of the mother-child dynamic.
Although secondary caregiver fathers generally expressed a desire to contribute, they stated being unaware of the responsibilities, taking a passive role, ``\textit{observed how mom did things}'' (P4D), and following the mother's lead, until more urgent situations occur in which they felt their participation was absolutely necessary. 

In addition, both fathers and mothers reported that mothers tend to have stronger connections to childcare resources, such as social media groups, friends, and family, which are often predominantly composed of other mothers. Fathers noted that being female facilitates these discussions for mothers within their social networks, while they, in contrast, found fewer informational resources specifically tailored to fathers. Additionally, some fathers expressed discomfort in participating in conversations that were predominantly female-oriented, such as asking about female daycare staff's contact or joining online mom groups, which made them feel awkward in these settings.

\subsubsection{Parental Collaboration Fosters Couple's Closeness.}
Despite the challenges, exhaustion, and emotional strain of being new parents, many couples experience a sense of closeness when witnessing their child’s development. The child’s growth provides a shared sense of accomplishment, as moments such as a child’s smile or milestones in their progress reinforce the perception that parenting is a collaborative and rewarding team effort. In discussing their shared parenting experiences, the P1 couple highlighted how witnessing their child's development together strengthens their bond. 
The couple illustrated the importance of these shared experiences in creating lasting memories, fostering closeness, and navigating the complexities of parenting together:
\begin{quote}
    [P1M]:\textit{``Parenting is tough, but having someone else there makes it all more rewarding.''}
    [P1D]: \textit{``When you both witness something happening with the baby and have the same reaction, it’s like a couple tasting food together and agreeing on the flavor. That shared experience creates memories, and you know you’ve gone through something significant together.''}
\end{quote}

Parents often developed a shared interest in their child's actions, whether positive or negative, as these experiences became common topics of conversation and connection between them. For couples who had recently become parents, working together to learn and perform basic caregiving tasks fostered a sense of closeness.
\begin{quote}
    \textit{``It's the feeling of learning these skills together, the emphasis being on ‘together’.''}-P8M
\end{quote}

While the child brought joy to the couple, the support they provided each other during this challenging period of their lives was also crucial, especially when one partner felt near a breaking point. Timely assistance from the other person not only offered a sense of relief but also strengthened their bond.
\begin{quote}
    \textit{``And also when we go through a difficult time together, it feels like you're on the brink of collapse, but the other person can step in and take over...In the middle of the night when he cried, I felt so overwhelmed, like I was about to lose it. Then my husband...would rush over when he heard me and ask, ‘What happened? Is everything okay?’ At that moment, it felt like I was being supported.''}-P7M
\end{quote}

\subsection{How do couples use technology in parental collaboration?}\label{How do couples use technology in parental collaboration}
Raising children is an intricate and dedicated process that often leads to not only physical fatigue but also emotional burnout. Throughout the interviews, couples mentioned diverse ways they used technology to facilitate the process of parenting and co-parenting. Some tools are specifically designed for parenting, such as pregnancy apps, baby monitoring cameras and apps, baby record-keeping apps to track routines, and daycare apps for parent-teacher communication. However, most participants adapted daily technology for their purposes, such as text messaging apps, to-do list apps, and photo-sharing apps. The following discussion will detail how current technologies used in co-parenting contributed to or fell short in roles such as information sharing and task coordination, memory sharing, enhancing communication, and emotional expression and exchange. The presentation of current practices of co-parenting technology use lays out the foundation to account for participants' co-parenting technology design.

\subsubsection{Sharing and Coordinating Tasks and Information are Essential}\label{Sharing and Coordinating Tasks and Information are Essential}

When children become the center of most parents' lives, parents use technology as a key tool in facilitating co-parenting. Our participants frequently mentioned that a significant aspect of using technology was sharing information about their children, which helped ensure smooth collaboration between co-parents and involved parties like relatives or caretakers. This approach minimized potential arguments and discrepancies while saving time and energy. 

Most couples in our study used baby activity tracking apps to help them keep track of their baby's daily routine, such as feeding, bowel movement, diaper change, etc. It provided them with a record of information that they could reference if there were any questions. 
Also, while they make use of current technology to share baby information, they wish it could better keep them on the same page. P5D shared how they used a baby app as a support to take turns to take care of the baby at night:
\begin{quote}
    \textit{``...when we started taking care of things separately, the information between the two of us would become unsynchronized. So, we needed an app to help us record everything that each person did at different times. This way, we can check the records... and we don't both need to be awake to exchange information.''} - P5D
\end{quote}
However, for some pairs of participants, only the mothers used baby activity tracking apps. The reasons why the couple did not use the apps collaboratively include 
(1) the app was for single-user use, 
(2) the couple did not think of using collaborative functions, 
(3) a parent did not find the usefulness of baby recording, 
and (4) the secondary caregiver relied on the primary caregiver for baby information retrieval, regardless of the sharing functions. 
Take P6 couple as an example, P6D relied on his wife to provide baby information, yet P6M hoped her husband could be more proactively involved in childcare.
\begin{quote}
    [P6D]: \textit{``The last time I used [global baby app] might be half a month ago... For information that requires action, my wife will either tell me directly or send me a message via LINE...and I'll respond, 'Okay, I have a task to do.'''}
    
    [P6M]: \textit{``...as part of the care team, I hope he could get a notification from the system. So I don't need to tell him orally.''}
\end{quote}
In this case, one parent relied on the other to know their baby's current status and could not proactively be involved in childcare. This situation could undermine couples' closeness as there was less interdependence within couples. Participants envisioned co-parenting technology that further supports couples' interdependence in Section \ref{How to better design co-parenting technology for closeness}.

\subsubsection{Communicating Logistics and the Lack of Emotion Awareness}
Couples often have to discuss things continuously about the baby or coordinate things in their lives. 
They can discuss in person or via messenger apps (e.g., LINE or Facebook Messenger). 
Despite that, some couples reported that communicating via instant messaging was time-consuming and prone to missing details, most couples among our participants used messenger apps to communicate with each other 
When it was not convenient to talk in person, such as when they were at work, or the partner was asleep. As P5D shared: 
\begin{quote}
    \textit{``Unless we're talking face-to-face, most of our discussions happen on LINE, especially since we're both working. So basically, we share much information about parenting, and when we encounter issues...''} - P5D
\end{quote}

While effective communication is crucial in both couple relationships and co-parenting practices, existing technology and app designs may not adequately support emotional expression. 
For instance, one participant, the primary caretaker of her child, relied on messaging apps to maintain the couple’s routine of saying goodnight. She recognized that \textit{``My husband sometimes felt disappointed (about this practice) because I was too tired to say goodnight to him in person''} (P6M). The husband expressed an insufficiency of technology in building an emotional connection, saying, 
\begin{quote}
    \textit{``I prefer face-to-face interaction, like saying goodnight, to increase the sense of closeness.''} - P6D
\end{quote}
Although current technology allows parents to communicate in a convenient and efficient way, closeness might be a trade-off. Participants envisioned potential functions of how co-parenting technology could balance efficiency and closeness in Section \ref{How to better design co-parenting technology for closeness}.

\subsubsection{Memorizing and Sharing Moments of Co-parental Experiences} \label{shared experiences}

Participants used various approaches to memorize and share co-parenting experiences. Some participants enjoyed learning about the baby’s development milestones together through a pregnancy application they use. Some participants used functions like creating a timeline of children’s growth by sorting baby photos and labeling dates, which helped parents review and appreciate their child’s growth and stay connected with the family.
As P7M noted, \textit{``If he was working that day or having a business trip, we wouldn’t stay in frequent contact. However, he could still see the photos I took that day through Google Photos.''}

Many couples choose to upload photos of their kids to ensure family members beyond the immediate household can jointly celebrate their kids' developmental journey. These apps often included features such as commenting and reacting, enabling relatives and parents to interact and express support. 

Some participants appropriated baby activity tracking apps, such as memorizing their efforts or knowing how their partner was doing, and feeling connected. As P5M shared,
\begin{quote}
    \textit{``...I never delete that breastfeeding app because it reminds me of my efforts. Pumping over 10,000 ml was a difficult and painful process for me, and I want to acknowledge that effort and thank myself.''} - P5M
\end{quote}

Some participants utilized technology to document and share meaningful moments related to raising their kids, thus fostering a sense of connection within the relationship. Some couples actively revisited past stories or data they have recorded in apps to relive those special memories, often evoking warmth and laughter in them. As P1M shared,
\begin{quote}
    \textit{``marriage is finding someone to witness…I jotted down those funny things in the app, and we would read them together and laugh about them.''} - P1M
\end{quote}
Yet, P1M further mentioned that the apps she used lacked functions that allowed her to document difficult moments. From these instances, current co-parenting technology offers functions to record babies' growth. However, participants needed to appropriate the existing functions to share their own experiences with their partners. Couples envisioned co-parenting technology that fosters them to participate in each other's parenting journey in Section \ref{How to better design co-parenting technology for closeness}.

\subsection{How to better design co-parenting technology for closeness?}\label{How to better design co-parenting technology for closeness}
Couples further shared ideas in envisioning activities about ways in which technology could better support them as co-parents and help them maintain a level of closeness in this process. We provided design prompts to guide these conversations. In this section, we present the insights we gleaned from the envisioning activities conducted with the parents.

Parents shared a greater desire for technology to facilitate better tasks, information sharing, and coordination for a smooth collaboration. They also pointed out that existing technology ignored the emotions they experienced in this life-changing journey of being new parents. They expressed the desire that technology could help commemorate and celebrate the key moments in their lives. Further, they believe that technology could help them maintain couples’ closeness during this challenging time in their lives by providing better support for their communication, emotional awareness, and emotional regulation.

\subsubsection{Information Sharing and Co-editing}\label{Information Sharing and Co-editing}

In co-parenting, childcare responsibilities are shared, which include not only taking care of the baby but also other household items that have to be taken care of at the same time. Participants offered ideas for improving technology to better support childcare responsibilities. For example, they believe that all baby-related information should be shared and allow co-editing, which can make it easier for both parents to be on the same page about child-related information and avoid belaboring one parent more than the other. One mom expressed the importance of information sharing as it could free her up when she wanted to take a break:
\begin{quote}
    \textit{``I record things (baby schedule) in the app, if I go out today and the dad stays home, I would hope that he also writes it down.''}-P4M
\end{quote}

However, parents struggled with baby information sharing, including not seeing the necessity of recording, and a lack of a collaborative editing function. 
\begin{quote}
    \textit{``For example, the time the baby goes to sleep, I also record it in the app, and it helps me calculate things like how long the baby has been sleeping, it's important for both parents. ...[But] this app doesn't allow for shared editing, so if both parents are raising the child together, the dad needs to be able to see it too, so it needs a collaborative editing feature.''} -P4M
\end{quote}

A detailed record of the baby's routine, which gives a more structured guide, could be useful for other family members who want to help with the baby, as it often feels “chaotic” when another person takes over. Such a guide could provide clear instructions for any caregiver, ensuring consistent care and attention to the child's needs.
\begin{quote}
  \textit{``I think of it more like the baby's instruction manual. For example, the baby's manual would include things like brushing their teeth after breakfast every day, drinking water first thing in the morning, applying sunscreen before going out if there’s going to be a lot of sun exposure, and making sure the diaper bag contains specific items when heading out. Essentially, if there were a baby manual, any caretaker would have clearer guidance on what to do next and what to be mindful of when caring for the child.''} -P7M
\end{quote}

Not only is sharing the baby’s schedule important, but sharing other child-related information, such as the baby's teacher and the baby's doctor. Making this information accessible to both parents could reduce the need for detailed labeling and memory recall. It could also help streamline communication and eliminate the burden of one parent having to provide contact details. 
One dad highlighted the importance of a shared contact feature due to the gender-based concern. This feature would be crucial and helpful in situations where their partner goes on a business trip. 
\begin{quote}
    \textit{``But this function is indeed necessary because... since the daycare teachers are all women, I don't feel comfortable asking for their phone numbers, so P1M handles that. This is a kind of gender-based division of labor...I think sharing contacts is very important. If, for some reason, my wife is on a business trip or unable to call the doctor, I might have the doctor's number, but I really don't have the teachers' phone numbers. So I think this feature would be helpful.''} - P1D
\end{quote}
P1M also agrees that this could be very helpful in avoiding unnecessary miscommunication between the couple.

\subsubsection{Foster Reliable and Satisfactory Task Sharing}\label{Foster Reliable and Satisfactory Task Sharing}

New parents often are not prepared for the long list of tasks they have to complete after the baby is born, which often require parents to redistribute their responsibilities. This process can be challenging for some couples. Some of our participants suggest that technology could help couples to better discuss task sharing, which would help mitigate conflicts down the road.

Participants discussed broadly how collaboration technology could support task arrangement.
While some parents hoped that the task-sharing app could automatically assign tasks to each parent because nobody needed to command task assignments, some suggested having an app that allows parents to assign tasks to individuals could be helpful in reducing conflicts between couples by promoting open communication and enabling couples to discuss and negotiate task exchanges to ensure the effective completion of responsibilities. 
\begin{quote}
   \textit{``Well, it's possible that the situation is that there’s no time, or maybe this part of the task isn’t being done well. At that point, I might say, "Are you having trouble completing that task? Do you want to swap tasks? … After we talk, we usually resolve things, right? We switch tasks after talking it over.''} -P8D
\end{quote}

When completing tasks together, it is common for couples to have different ideas. Some couples suggest that technology can help them mitigate disagreements and reach consensus. P4M recalled a conflict that happened between her and her husband about a household chore. She called this “the most memorable disagreement”. They had different perspectives on when to wash dishes. The father preferred waiting, while the mother, concerned about time constraints before school, wanted to clean them immediately. However, due to the division of labor at the time, with the mother breastfeeding, the father handled more of the dishwashing, which occasionally led to disagreements.
Therefore, P4M thought that it would be beneficial to agree upon a deadline: 
\begin{quote}
    \textit{``Yeah, and if we were to systematize it, we might have to be very specific, like saying Dad should finish washing by a certain time, assigning tasks with deadlines.''} -P4M
\end{quote}
P4D agrees that even though this approach may appear “harsh” or “in-personable”, it gets things done and reduces conflicts: 
\begin{quote}
    \textit{``I would say that, by turning everything into rules, it might feel a bit impersonal. But if we list everything out and follow the schedule, it could reduce potential issues between us. It’s subtle; it might make us feel less close, but it could help solve problems and avoid creating a bad atmosphere at home.''} -P4D
\end{quote}
In this case, P4 couple's design aimed to avoid conflicts that make them distant. Although this kind of design could take away couple interactions, such as communicating each other's values, the P4 couple chose to avoid conflicts by setting clear boundaries at the cost of being less close.

\subsubsection{Awareness of Mental State Elicit Empathy and Proactive Support}\label{Awareness of Mental State Elicit Empathy and Proactive Support}

Parents envisioned various technological support to know their partners' mental state because knowing their partners' situations could enable empathy. Empathy is a key consideration when collaborating with their partners in determining whether they would provide support or take over tasks. For example, P5D noted that emotional capacity often sets the limits for parenting responsibilities.
\begin{quote}
    \textit{``...no matter how much you discuss, you can't really quantify the amount of effort or hardship. In the end, it comes down to the emotional aspect—whether one person's mental energy can handle that many responsibilities.''} -P5D
\end{quote}

Parents experienced various emotional hardships, such as sleep deprivation, work stress, loss of family members, and postpartum depression. 
Postpartum depression was commonly mentioned by our participants.
Some fathers sometimes felt they could not completely empathize with their wives because they did not personally experience it. 
Many conflicts were due to the lack of empathy for their partners' emotional struggles.
Therefore, several solutions were proposed by participants to be aware of their partners' emotional state, such as manual depression tracking with a daily depression questionnaire, self-reporting of emotions, or an emotion communication enhancement card that a parent could send to their partner and guide their partners to share emotions.
Parents also consider functions that could predict emotions, such as based on auto-detection of facial expressions or sleep. In this way, they could support their partner before it was too late. For example, both P10M and P10D shared the same view that they wanted a shared emotion report.  
\begin{quote}
    \textit{``It's something that both of us can see. For example, like my wife just mentioned, she hopes for a postpartum depression questionnaire to pop up automatically.
    ...If he sees that today's score is, say, five or a bit high, then he should be more aware and considerate of his wife today. ... Vice versa...the other person [his wife] sees it, comes over, and says something to comfort you. I think that's a huge encouragement. Yes, just having someone empathize with you is already a big help.''}-P10D 
\end{quote}
In this case, the awareness of emotions could lead to acknowledgment of these struggles and subsequent supportive actions from the partner. 

Furthermore, many couples also expressed that not only their partners but also themselves could benefit from knowing their own emotions, as they could be accumulating negative emotions without knowing, and a reality check could help them to recognize it and take some action. Participants proposed various interventions regarding action taking, including emotional self-regulation guidance for parenting pain or a compassionate notification suggesting taking a rest or asking for help. 
Participants also considered the situation when both parents were near their emotional limits. P3D suggested a function to prompt the couple to seek external help, such as the grandparents.

While emotional regulation and support are crucial for couples in their early parenting experience, participants also showed interest in using technology to smooth communication with each other about their negative emotions, aiming to mitigate potential conflicts. Functions participants proposed include an instant messaging app that transforms the direct message to a subtle and less harsh message and a function providing communication templates to help refine messages to be thoughtful and considerate, avoiding emotionally charged or commanding language, which can lead to misunderstanding and hurt feelings between the couple.
\begin{quote}
    \textit{``Or provide some communication templates...like when I want to say something, if I say it in a very emotional or commanding way, without considering the other person's perspective, it can lead to both sides getting hurt. But if I have a better way of saying it, or if I type out a message and it helps me refine it, maybe that could work...''}-P4M
\end{quote}

\subsubsection{Foster Positive Emotions and Fun Shared Experiences}\label{D: positive emotion}
All parents we interviewed use technology to support their co-parenting. They pointed out that current technology mostly focuses on tracking and supporting tasks, such as to-do lists, baby scheduling apps, and breastfeeding trackers, but they ignore the positive emotions parents experience when sharing or completing these tasks.

Parenting tasks are not simply tasks, they are records of their parenting journey. For completed tasks and experiences, a “done list” could help commemorate the parenting journey and the emotions behind this life-changing experience and foster a sense of gratitude towards both themselves and their partner. 
\begin{quote}
   \textit{``
   It's like how I never delete that breastfeeding app because it reminds me of my efforts... I think that on the journey of parenting, we often forget the great things we've done or that our partner has done...
   It’s not about it being a labor-intensive record; it’s about the emotional desire to hold onto that memory.''}-P5M
\end{quote}
Our participants not only wanted to feel gratitude but also express it. As P5 couple proposed,
\begin{quote}
   \textit{``I think an achievement system would be good... Even if it's just adding one more task to daily life, I think it helps the relationship. Like giving a compliment to the other person...''} - P5M
\end{quote}
A design that nudges couples to express gratitude could be helpful for closeness. In addition, P5D also believed that a "done list" could help him and his wife to have empathy for each other: \textit{``I imagine it could help both of us understand each other's contributions better…It might increase empathy.''}

Some parents felt a sense of closeness and accomplishment when they completed something together. Functions parents suggested include prompting parents to share the completion with each other, or prompting parents to show gratitude towards each other for tasks that are completed, or tasks they may do additionally for each other.
\begin{quote}
    \textit{`` 
    It’s like a sense of accomplishment as parents—we’ve made it through another day…[when we complete everything for the day] we can celebrate it, or maybe finish early and have some personal time left for ourselves.''} - P8D
\end{quote}
P8D further mentioned that current technology (e.g., calendar or to-do list) does not help to capture that.

One couple discussed that visualizing the fact that two parents were parenting together could reignite their passion for building a family together. In the scenario that a couple had to split the tasks to complete the childcare tasks, one couple thought that visualization of togetherness could serve as a reminder that they were working together instead of parenting alone.
\begin{quote}
    \textit{``We’ve always assumed that as a couple, we would take care of our child together. I believe both of us have been working hard on this shared foundation of understanding, but this foundation is so important that it might have become invisible to us. However, it is not bad to bring it out and review it. It could reinvigorate us even more.''} - P6D
\end{quote}

In addition to eliciting positive emotions from the parental collaboration, technology could potentially suggest something fun for couples to be more intimate, like a "date invitation" or suggesting topics for conversations:
\begin{quote}
    \textit{``Randomly throwing out a topic or something like that, but the topic can't be too complicated. If it's too complex, it becomes a task that you need to complete.''} -P6M
\end{quote}

Another function some participants suggested is that the applications could be used to express personal intimate needs, such as wanting to spend time together, in a more engaging and playful way, like sending an invitation for an activity. This approach is perceived as more fun and creative than direct communication methods.
\begin{quote}
    \textit{``you can express emotional needs, like if you really want to go out, or if the two of you want to do something alone together, like watching a movie or going to a night market. You could use this app to send an invitation in a way that’s more fun or cute, rather than just saying it directly through LINE or Messenger.''} - P7M
\end{quote}

\section{Discussion}
Our study focuses on examining ways in which couples use current technology to support collaborations and closeness. We found that, other than a few technologies that were designed for parenting, most parents appropriated technology for parenting purposes. We also discovered that current technology design ignores emotions in the parenting experience. Through our design sessions, we collaborated with new parent couples and revealed opportunities for co-parenting technology to increase new parent couples' closeness through integrating positive emotions into parenting and fostering interdependence.

\subsection{Positive emotions in co-parenting increase closeness}
\subsubsection{Collaboration enables positive emotions.}
Prior parental collaboration work found that mothers hope to bond with their support network through having more family time to know one another more and show care and appreciation~\cite{ko2022mobilizing}. By having a couple as a unit to envision instead of focusing on the mothers, our work extends prior work by revealing that not only mothers but both parents desire to bond and cultivate joy, togetherness, and understanding. Despite the bonding needs, our findings also indicate that current technologies for supporting the process of parenting are limited in their capacity to foster these positive emotions among couples. As illustrated in Section \ref{shared experiences}, although parents use technologies to share and enjoy their baby's development through baby photo apps and pregnancy apps to foster interrelationships bonds between themselves (resonates with prior work~\cite{lu2024unpacking, lu2024examining}, some parents still appropriate the current baby task records to help with the process of sense-making about their parenthood experiences from the perspective of a couple and individually (e.g., attaching sentiments value to the task). This understanding justifies why our participants proposed various designs to enrich positive emotions as detailed in Section \ref{D: positive emotion}. 
In addition, prior work only described the need for bonding; our results in Section \ref{D: positive emotion} flesh out practical approaches and opportunities for how technology can support couples in fulfilling the bonding needs in hectic parental collaboration scenarios. 

One silent trait of participants' designs for promoting positive emotions is to use collaboration and efforts as an opportunity for fostering positive emotions. The reason why parents use collaboration efforts for positive emotions might be that intense collaboration creates a deep sense of bonding through shared experiences~\cite{rosenfeld2023resilience} and appreciation~\cite{white2024father}. 
We use three example design opportunities to demonstrate how to incorporate collaboration efforts for positive emotions:
\begin{itemize}
    \item \textbf{Celebrate collaborative achievement.} 
    Prior work suggests parents celebrate their similarities~\cite{yarosh2016best}, we think it is also critical to celebrate their collaborative work and validate each other. 
    Our participants desire to have technology that nudges them to celebrate their achievements and validate each other's progress. For example, a "done list" of childcare tasks can serve as a record of achievement and could nudge parents to celebrate the moment when they complete all the tasks with some intimate interaction, such as hugging each other or taking a family photo together. 

    \item \textbf{Surface shared experiences.} 
    In light of the desire for couples to have shared experiences while considering all the childcare workloads that require parents to work separately, we see the opportunity to design technology to elicit shared experiences from their separate tasks. 
    Prior work shows that a display at home juxtaposing family members' current location together could not only have practical utilization of knowing where the families are but also enhance the sense of togetherness~\cite{brown2007locating}. 
    Technology could promote a sense of doing activities together when it is easier for family members to do the shared activities separately~\cite{christensen2019together}, 
    a visualization that juxtaposes the work each of them has done, is doing, and sharing and responding how they feel, has the potential to facilitate shared experience. For example, when parents complete a task, it can be represented as adding a piece to a jigsaw puzzle—one that they are collaboratively assembling together. Then, each parent could share their thoughts on each piece. The shared memory, therefore, includes not only the baby's growth but also the couple's shared parenting experiences.

    \item \textbf{Trigger and express acknowledgment and appreciation}
    Parental collaboration involves efforts, contributions, and support. These are great sources of appreciation and admiration for the partner. One's sense of gratitude toward their partner predicts both their own and their partner's long-term relationship satisfaction~\cite{gordon2011have}. Prior work suggests that tracking the husband's childcare efforts could trigger the wife's awareness of the husband's efforts and the appreciation for their husband~\cite{jo2020understanding}. Our findings support prior work, as our participants also proposed to use a done list of childcare tasks to memorize their partners' contributions.
    In addition, feeling appreciated catalyzes relationship-maintaining behaviors that signal responsiveness and commitment, fostering reciprocal gratitude and initiating a self-reinforcing cycle of relational enhancement~\cite{algoe2012find, gordon2012have}. The done list could further nudge the couple to express gratitude for enhancing the positive responsiveness-gratitude cycle.
\end{itemize}

We would like to note that fostering positive emotions in stressful collaborative contexts can be complex. For instance, making efforts and achievements more visible may also expose disparities in contribution, potentially triggering conflict. This highlights the need for further design considerations to support parents in navigating and discussing effort imbalances, especially given that each parent may assign different meanings and values to specific tasks. While system-prompted appreciation might encourage the expression of positive emotions, it also carries the risk of being perceived as inauthentic. We encourage future research to further explore these trade-offs and develop designs that minimize potential negative consequences.

\subsection{Interdependent co-parenting increases closeness}
Prior work has explored technologies that support couples' closeness through the perspectives of (a) intimacy~\cite{vetere2005mediating}, (b) shared experiences (e.g., collaboration experiences~\cite{he2013couple}), and (c) self-disclosure (e.g., expression of desires~\cite{ellegaard2021shaping}).
Our findings extend this line of research by highlighting how parents desire not only the described closeness but also access to technology that will facilitate interdependence, such as the passive parent's initiative in taking over child-care-related tasks, increasing parental competence for reliance, and having childcare consensus with mutual respect.

\subsubsection{Empathy Enables Action: From reactive to proactive participation and takeover} \label{Empathy Enables Action: From reactive to proactive participation and takeover}

Existing literature has shown the possibility for empathy and social support to be triggered by tracking the mother through passive and manual stress tracking~\cite{song2020bodeum}, tracking of childcare time~\cite{jo2020understanding}, and pregnancy tracking~\cite{lu2024examining}. 
In Section \ref{Awareness of Mental State Elicit Empathy and Proactive Support}, parents proposed designs to not only track mothers but also both parents to mutually understand mental states (e.g., depression, emotion, and sleep) within the couple and thus enable the provision of support to each other. Reflecting on our findings, we argue that technology for co-parenting could support parents to expand their awareness of each other's well-being status through mutual multi-factor tracking (e.g., childcare time, stress, emotion, and sleep) so that parents could understand each other's situation and needs and provide support. 
This approach is aligned with Reis's intimacy process model~\cite{reis2018intimacy} that self-disclosure and the corresponding partner response could increase closeness. When a parent self-discloses his or her mental state and their partner responds with care, they become closer to each other. In addition, we believe mutuality is grounded in the concept of a co-parenting context, where both parents are regarded as equal stakeholders~\cite{ramchandani2013early, park2023actor} and mutual awareness is key to successful collaboration~\cite{dourish1992awareness}. 
To illustrate, when the couple described their vision of how technology fosters interdependence in parenting, they envisioned that their partner would become more proactive in taking on tasks at home and childcare-related. Participants elaborated on how technology could foster interdependence in co-parenting by learning insights about, for example, whether the partner's sleep quality and emotions are good during the stage of postpartum recovery. This understanding further revealed that information such as each parent's well-being promotes a sense of self-awareness within new parents. When parents are aware of their partner's and their own status, it is easier for them to become more proactive in situations when a parent experiences emotional hardships, such as sleep deprivation or postpartum depression, and thus facilitates interdependence.

\subsubsection{Coordinating parenting information and values for reciprocal reliance and respect} \label{Coordinating parenting information and values for reciprocal reliance}
As detailed in Section \ref{Distant Moments in Parental Collaboration}, several fathers reported feeling less helpful and more passive in their parenting roles, often following their wives' lead. Some mothers expressed concerns about their spouses' parenting abilities and wished for proactive involvement. Both parents experienced feelings of distance; mothers felt they were handling parenting responsibilities alone, while fathers felt guilty and impotent.
This finding aligns with~\cite{ko2022mobilizing, lu2024unpacking}.\avon{this work extends...}
This dynamic may be influenced by societal expectations, which traditionally view mothers as the primary caregivers and fathers as secondary caregivers or providers~\cite{hsiu2011parent}. 
For example, we found gendered information practices in our data. Participants reported that mothers retrieve childcare resources more easily and actively seek parenting information. 
This phenomenon is also observed by researchers who studied healthcare services~\cite{darroch2023m}, pregnancy apps~\cite{thomas2018appy, lu2024examining}, and childcare apps~\cite{jensen2019into}. For example, \citet{thomas2018appy} discovered that parenting information for fathers appears overly simplistic, portraying fathers as secondary givers and incompetent or less involved in parenting compared to mothers. 
Moreover, some of our mother participants took more childcare tasks because of the parental competence differences between the two parents.
Prior work shows that mothers may develop gatekeeping behaviors, such as feeling it easier to handle the tasks by themselves or manage tasks, due to worries about fathers' parenting quality~\cite{tu2014study, allen1999maternal}. Yet, previous research also indicates that these behaviors could unintentionally inhibit fathers from more parental involvement~\cite{tu2014study, diniz2021father}.

These dynamics can hinder closeness from the perspective of interdependence because parents do not interdepend on each other if one of them is less competent or if parents do not have a consensus on how to do childcare tasks.
Reflecting on the need for interdependence, participants proposed various ways to align parents' childcare ability and standards, as detailed in Section \ref{Information Sharing and Co-editing} and \ref{Foster Reliable and Satisfactory Task Sharing}. 
We believe it is critical to see parents as equal stakeholders in the co-parenting context. Technology should be designed for both parents and see them as a family unit. It could provide an infrastructure that facilitates proactive participation and enables coordinating parenting information and values, thus serving as a start to empower fathers and enable interdependence between both parents. 
We discuss design considerations that foster closeness by enabling co-parents' interdependence in parental collaboration within this context. These include facilitating reliance, enabling the takeover of tasks, and fostering mutual respect between a couple.

\paragraph{Tailored parenting knowledge base of childcare tasks.} 
An increase in parenting knowledge could empower less competent parents. According to our findings that parents proposed designs for sharing parenting knowledge, we believe that more skillful parents in parenting are willing to share their parenting knowledge, and less skillful parents are willing to learn. 
A knowledge base could make the complex childcare tasks more visible by including details of childcare knowledge and a list of tasks that co-parents can reference. Just-in-time information provision could also be used to scaffold the learning of childcare knowledge. 
For passive parents, the current way they gained parenting knowledge relied on their partners to the extent that their partners felt burdened. Such a knowledge base could nudge and empower passive parents to gain parental competence. In addition, some parents are more detail-oriented, and some are more laid-back. We think a knowledge base of childcare tasks can serve as a starting point for parents to discuss how they want to take care of the baby. The default knowledge base could be integrated into a pregnancy app to help both parents prepare for early parenthood with realistic, research-based guidance from family studies on parenting and co-parenting ~\cite {entsieh2016first, eira2021co}. 
This is also aligned with the previous HCI literature suggesting to prepare parents for the chaos caused by an urgent scenario~\cite{nikkhah2022family}

Moreover, there could be a discussion button to nudge parents to discuss when a parent feels a discussion is needed.
Thus, couples could discuss and tailor the knowledge base to meet their childcare styles.

\section{Limitation and future work}
There were several sources of limitations for this work. Regarding the participant demographics, potentially affected by social media recruitment, all our participants were married, aged 30 and above, with medium to high socioeconomic status. None of the fathers was the sole primary caregiver. Family study research has shown that economic strain was linked to parents' lower relationship quality~\cite{broderick2019interparental}. Therefore, financial factors may warrant additional consideration when designing for relational closeness for low-income families. Moreover, all our participants were from Taiwan, where paternal leave is less common than in many Western countries like the US. Cultural norms shape how parents view closeness, collaboration, and co-parenting technologies~\cite{pfeifer2013perceived, newland2013parent}. In Taiwanese contexts, design for interdependence could thus focus on helping fathers catch up on childcare. By contrast, cultures with more common paternal leave might support different models of interdependence, such as the redistribution of childcare tasks based on child development~\cite{lin2024ultimately}. All participants were heterosexual, and we believe same-sex couples have different family dynamics that would affect the findings. Also, all couples live together, and the need for closeness could be different from couples who live apart. We encourage future work to broaden the investigation to include various populations, such as young parents, low-income couples, LGBTQ+ couples, unmarried couples, couples living apart, or couples from other countries. 

Regarding the study method, we had the couples co-design together to reach a consensus on their proposed designs. Yet, participants might hesitate to share some thoughts that upset their partners. In future studies, we suggest having some couples do separate design sessions and some couples design technology together to compensate for the disadvantages of the two approaches. 
In addition, as an exploratory study with a limited number of participants, we do not intend to draw conclusions and generalize from the findings. Yet, the qualitative insights derived from this study will contribute to further discussion about the design of technology for closeness and parental collaboration. 

Last, we found that grandparents and extended family play a critical role in couples' closeness and co-parenting practices. In East Asian culture, child-rearing could be an effort of a bigger group of people other than the couple. Thus, future work could broaden the range of investigation to extended families.

\section{Conclusion}
Couples often experience a decline in closeness as they navigate the demands of parenthood. While existing technologies have facilitated parenting tasks and collaboration, they fall short in fostering closeness between co-parents. To explore how technology could better support closeness, we conducted initial sessions of co-design workshops with 10 new parent couples. We examined their current use of technology for parental collaboration and how they adapted both current and emerging tools to foster closeness, and we found that parents proposed extensive design ideas related to various aspects of closeness. We identified opportunities for technology to further enhance closeness in co-parenting, including promoting interdependence and integrating positive emotions into parenting.

\begin{acks}
We thank our participants for generously sharing their invaluable experiences and thoughts with us. We also thank the anonymous reviewers and our friends and colleagues (Dakuo Wang, Catherine Wieczorek-Berkes, and Sarah Tran) for their insightful suggestions and comments.
\end{acks}

\bibliographystyle{ACM-Reference-Format}
\bibliography{sample-base}


\begin{thebibliography}{97}


\ifx \showCODEN    \undefined \def \showCODEN     #1{\unskip}     \fi
\ifx \showDOI      \undefined \def \showDOI       #1{#1}\fi
\ifx \showISBNx    \undefined \def \showISBNx     #1{\unskip}     \fi
\ifx \showISBNxiii \undefined \def \showISBNxiii  #1{\unskip}     \fi
\ifx \showISSN     \undefined \def \showISSN      #1{\unskip}     \fi
\ifx \showLCCN     \undefined \def \showLCCN      #1{\unskip}     \fi
\ifx \shownote     \undefined \def \shownote      #1{#1}          \fi
\ifx \showarticletitle \undefined \def \showarticletitle #1{#1}   \fi
\ifx \showURL      \undefined \def \showURL       {\relax}        \fi
\providecommand\bibfield[2]{#2}
\providecommand\bibinfo[2]{#2}
\providecommand\natexlab[1]{#1}
\providecommand\showeprint[2][]{arXiv:#2}

\bibitem[Abreu-Afonso et~al\mbox{.}(2022)]%
        {abreu2022couple}
\bibfield{author}{\bibinfo{person}{Jos{\'e} Abreu-Afonso}, \bibinfo{person}{Maria~Meireles Ramos}, \bibinfo{person}{Ineˆs Queiroz-Garcia}, {and} \bibinfo{person}{Isabel Leal}.} \bibinfo{year}{2022}\natexlab{}.
\newblock \showarticletitle{How couple’s relationship lasts over time? A model for marital satisfaction}.
\newblock \bibinfo{journal}{\emph{Psychological reports}} \bibinfo{volume}{125}, \bibinfo{number}{3} (\bibinfo{year}{2022}), \bibinfo{pages}{1601--1627}.
\newblock


\bibitem[Algoe(2012)]%
        {algoe2012find}
\bibfield{author}{\bibinfo{person}{Sara~B Algoe}.} \bibinfo{year}{2012}\natexlab{}.
\newblock \showarticletitle{Find, remind, and bind: The functions of gratitude in everyday relationships}.
\newblock \bibinfo{journal}{\emph{Social and personality psychology compass}} \bibinfo{volume}{6}, \bibinfo{number}{6} (\bibinfo{year}{2012}), \bibinfo{pages}{455--469}.
\newblock


\bibitem[Allen and Hawkins(1999)]%
        {allen1999maternal}
\bibfield{author}{\bibinfo{person}{Sarah~M Allen} {and} \bibinfo{person}{Alan~J Hawkins}.} \bibinfo{year}{1999}\natexlab{}.
\newblock \showarticletitle{Maternal gatekeeping: Mothers' beliefs and behaviors that inhibit greater father involvement in family work}.
\newblock \bibinfo{journal}{\emph{Journal of Marriage and the Family}} (\bibinfo{year}{1999}), \bibinfo{pages}{199--212}.
\newblock


\bibitem[Aron and Aron(1997)]%
        {aron1997self}
\bibfield{author}{\bibinfo{person}{Arthur Aron} {and} \bibinfo{person}{Elaine~N Aron}.} \bibinfo{year}{1997}\natexlab{}.
\newblock \showarticletitle{Self-expansion motivation and including other in the self.}
\newblock  (\bibinfo{year}{1997}).
\newblock


\bibitem[Banks(1998)]%
        {banks1998lives}
\bibfield{author}{\bibinfo{person}{James~A Banks}.} \bibinfo{year}{1998}\natexlab{}.
\newblock \showarticletitle{The lives and values of researchers: Implications for educating citizens in a multicultural society}.
\newblock \bibinfo{journal}{\emph{Educational researcher}} \bibinfo{volume}{27}, \bibinfo{number}{7} (\bibinfo{year}{1998}), \bibinfo{pages}{4--17}.
\newblock


\bibitem[B{\'e}cotte et~al\mbox{.}(2023)]%
        {becotte2023positive}
\bibfield{author}{\bibinfo{person}{Katherine B{\'e}cotte}, \bibinfo{person}{Audrey Brassard}, \bibinfo{person}{Anne Brault-Labb{\'e}}, \bibinfo{person}{Anne-Laurence Gagn{\'e}}, {and} \bibinfo{person}{Katherine P{\'e}loquin}.} \bibinfo{year}{2023}\natexlab{}.
\newblock \showarticletitle{Positive relationship adaptation of couples transitioning to parenthood: An interpretative phenomenological analysis}.
\newblock \bibinfo{journal}{\emph{Family Relations}} \bibinfo{volume}{72}, \bibinfo{number}{4} (\bibinfo{year}{2023}), \bibinfo{pages}{2251--2269}.
\newblock


\bibitem[Berscheid et~al\mbox{.}(1989)]%
        {berscheid1989relationship}
\bibfield{author}{\bibinfo{person}{Ellen Berscheid}, \bibinfo{person}{Mark Snyder}, {and} \bibinfo{person}{Allen~M Omoto}.} \bibinfo{year}{1989}\natexlab{}.
\newblock \showarticletitle{The Relationship Closeness Inventory: Assessing the closeness of interpersonal relationships.}
\newblock \bibinfo{journal}{\emph{Journal of personality and Social Psychology}} \bibinfo{volume}{57}, \bibinfo{number}{5} (\bibinfo{year}{1989}), \bibinfo{pages}{792}.
\newblock


\bibitem[Borgkvist et~al\mbox{.}(2020)]%
        {borgkvist2020unfortunately}
\bibfield{author}{\bibinfo{person}{Ashlee Borgkvist}, \bibinfo{person}{Jaklin Eliott}, \bibinfo{person}{Shona Crabb}, {and} \bibinfo{person}{Vivienne Moore}.} \bibinfo{year}{2020}\natexlab{}.
\newblock \showarticletitle{“Unfortunately I’ma massively heavy sleeper”: An analysis of fathers’ constructions of parenting}.
\newblock \bibinfo{journal}{\emph{Men and Masculinities}} \bibinfo{volume}{23}, \bibinfo{number}{3-4} (\bibinfo{year}{2020}), \bibinfo{pages}{680--701}.
\newblock


\bibitem[Branham and Harrison(2013)]%
        {branham2013designing}
\bibfield{author}{\bibinfo{person}{Stacy Branham} {and} \bibinfo{person}{Steve Harrison}.} \bibinfo{year}{2013}\natexlab{}.
\newblock \showarticletitle{Designing for collocated couples}.
\newblock \bibinfo{journal}{\emph{Connecting families: the impact of new communication technologies on domestic life}} (\bibinfo{year}{2013}), \bibinfo{pages}{15--36}.
\newblock


\bibitem[Broderick et~al\mbox{.}(2019)]%
        {broderick2019interparental}
\bibfield{author}{\bibinfo{person}{Amanda~V Broderick}, \bibinfo{person}{Gina~M Brelsford}, {and} \bibinfo{person}{Martha~E Wadsworth}.} \bibinfo{year}{2019}\natexlab{}.
\newblock \showarticletitle{Interparental relationships among low income, ethnically diverse, two-parent cohabiting families}.
\newblock \bibinfo{journal}{\emph{Journal of Child and Family Studies}}  \bibinfo{volume}{28} (\bibinfo{year}{2019}), \bibinfo{pages}{2259--2271}.
\newblock


\bibitem[Brown et~al\mbox{.}(2007)]%
        {brown2007locating}
\bibfield{author}{\bibinfo{person}{Barry Brown}, \bibinfo{person}{Alex~S Taylor}, \bibinfo{person}{Shahram Izadi}, \bibinfo{person}{Abigail Sellen}, \bibinfo{person}{Joseph~Jofish’ Kaye}, {and} \bibinfo{person}{Rachel Eardley}.} \bibinfo{year}{2007}\natexlab{}.
\newblock \showarticletitle{Locating Family Values: A Field Trial of the Whereabouts Clock: (Nominated for the Best Paper Award)}. In \bibinfo{booktitle}{\emph{UbiComp 2007: Ubiquitous Computing: 9th International Conference, UbiComp 2007, Innsbruck, Austria, September 16-19, 2007. Proceedings 9}}. Springer, \bibinfo{pages}{354--371}.
\newblock


\bibitem[Campbell(2023)]%
        {campbell2023two}
\bibfield{author}{\bibinfo{person}{Cynthia~G Campbell}.} \bibinfo{year}{2023}\natexlab{}.
\newblock \showarticletitle{Two decades of coparenting research: a scoping review}.
\newblock \bibinfo{journal}{\emph{Marriage \& Family Review}} \bibinfo{volume}{59}, \bibinfo{number}{6} (\bibinfo{year}{2023}), \bibinfo{pages}{379--411}.
\newblock


\bibitem[Carroll(2003)]%
        {carroll2003scenario}
\bibfield{author}{\bibinfo{person}{John~M Carroll}.} \bibinfo{year}{2003}\natexlab{}.
\newblock \bibinfo{booktitle}{\emph{Scenario-based design}}.
\newblock \bibinfo{publisher}{MIT Press}.
\newblock


\bibitem[Christensen et~al\mbox{.}(2019)]%
        {christensen2019together}
\bibfield{author}{\bibinfo{person}{Peter~Kn{\o}sgaard Christensen}, \bibinfo{person}{Christoffer~{\O}land Skovgaard}, {and} \bibinfo{person}{Marianne~Graves Petersen}.} \bibinfo{year}{2019}\natexlab{}.
\newblock \showarticletitle{Together together: Combining shared and separate activities in designing technology for family life}. In \bibinfo{booktitle}{\emph{Proceedings of the 18th acm international conference on interaction design and children}}. \bibinfo{pages}{374--385}.
\newblock


\bibitem[Christopher et~al\mbox{.}(2015)]%
        {christopher2015marital}
\bibfield{author}{\bibinfo{person}{Caroline Christopher}, \bibinfo{person}{Tomo Umemura}, \bibinfo{person}{Tanya Mann}, \bibinfo{person}{Deborah Jacobvitz}, {and} \bibinfo{person}{Nancy Hazen}.} \bibinfo{year}{2015}\natexlab{}.
\newblock \showarticletitle{Marital quality over the transition to parenthood as a predictor of coparenting}.
\newblock \bibinfo{journal}{\emph{Journal of Child and Family Studies}} \bibinfo{volume}{24}, \bibinfo{number}{12} (\bibinfo{year}{2015}), \bibinfo{pages}{3636--3651}.
\newblock


\bibitem[Cowan and Cowan(1992)]%
        {cowan1992partners}
\bibfield{author}{\bibinfo{person}{Carolyn~Pape Cowan} {and} \bibinfo{person}{Philip~A Cowan}.} \bibinfo{year}{1992}\natexlab{}.
\newblock \bibinfo{booktitle}{\emph{When partners become parents: The big life change for couples.}}
\newblock \bibinfo{publisher}{basic books}.
\newblock


\bibitem[Cox and Paley(1997)]%
        {cox1997families}
\bibfield{author}{\bibinfo{person}{Martha~J Cox} {and} \bibinfo{person}{Blair Paley}.} \bibinfo{year}{1997}\natexlab{}.
\newblock \showarticletitle{Families as systems}.
\newblock \bibinfo{journal}{\emph{Annual review of psychology}} \bibinfo{volume}{48}, \bibinfo{number}{1} (\bibinfo{year}{1997}), \bibinfo{pages}{243--267}.
\newblock


\bibitem[Darroch et~al\mbox{.}(2023)]%
        {darroch2023m}
\bibfield{author}{\bibinfo{person}{Francine~E Darroch}, \bibinfo{person}{John~L Oliffe}, \bibinfo{person}{Gabriela Gonzalez~Montaner}, {and} \bibinfo{person}{Jessica~M Webb}.} \bibinfo{year}{2023}\natexlab{}.
\newblock \showarticletitle{“I’m Trying to be There for my Kids”: A Needs Analysis of Fathers Who Experience Health Inequities in Vancouver, Canada}.
\newblock \bibinfo{journal}{\emph{Men and Masculinities}} \bibinfo{volume}{26}, \bibinfo{number}{2} (\bibinfo{year}{2023}), \bibinfo{pages}{229--250}.
\newblock


\bibitem[Davidoff et~al\mbox{.}(2010)]%
        {davidoff2010routine}
\bibfield{author}{\bibinfo{person}{Scott Davidoff}, \bibinfo{person}{John Zimmerman}, {and} \bibinfo{person}{Anind~K Dey}.} \bibinfo{year}{2010}\natexlab{}.
\newblock \showarticletitle{How routine learners can support family coordination}. In \bibinfo{booktitle}{\emph{Proceedings of the SIGCHI Conference on Human Factors in Computing Systems}}. \bibinfo{pages}{2461--2470}.
\newblock


\bibitem[Deci and Ryan(2014)]%
        {deci2014autonomy}
\bibfield{author}{\bibinfo{person}{Edward~L Deci} {and} \bibinfo{person}{Richard~M Ryan}.} \bibinfo{year}{2014}\natexlab{}.
\newblock \showarticletitle{Autonomy and need satisfaction in close relationships: Relationships motivation theory}.
\newblock \bibinfo{journal}{\emph{Human motivation and interpersonal relationships: Theory, research, and applications}} (\bibinfo{year}{2014}), \bibinfo{pages}{53--73}.
\newblock


\bibitem[Derix and Leong(2020)]%
        {derix2020probes}
\bibfield{author}{\bibinfo{person}{Eleanor~Chin Derix} {and} \bibinfo{person}{Tuck~Wah Leong}.} \bibinfo{year}{2020}\natexlab{}.
\newblock \showarticletitle{Probes to explore the individual perspectives on technology use that exist within sets of parents}. In \bibinfo{booktitle}{\emph{Proceedings of the 2020 ACM Designing Interactive Systems Conference}}. \bibinfo{pages}{519--531}.
\newblock


\bibitem[Deuff et~al\mbox{.}(2022)]%
        {deuff2022together}
\bibfield{author}{\bibinfo{person}{Dominique Deuff}, \bibinfo{person}{Isabelle Milleville-Pennel}, \bibinfo{person}{Ioana Ocnarescu}, \bibinfo{person}{Dora Garcin}, \bibinfo{person}{Corentin Aznar}, \bibinfo{person}{Sim{\'e}on Capy}, \bibinfo{person}{Shohei Hagane}, \bibinfo{person}{Pablo~Felipe Osorio~Marin}, \bibinfo{person}{Enrique Coronado~Zuniga}, \bibinfo{person}{Liz Rincon~Ardila}, {et~al\mbox{.}}} \bibinfo{year}{2022}\natexlab{}.
\newblock \showarticletitle{Together alone, Y{\=o}kobo, a sensible presence robject for the home of newly retired couples}. In \bibinfo{booktitle}{\emph{Proceedings of the 2022 ACM Designing Interactive Systems Conference}}. \bibinfo{pages}{1773--1787}.
\newblock


\bibitem[Diniz et~al\mbox{.}(2021)]%
        {diniz2021father}
\bibfield{author}{\bibinfo{person}{Eva Diniz}, \bibinfo{person}{T{\^a}nia Brand{\~a}o}, \bibinfo{person}{L{\'\i}gia Monteiro}, {and} \bibinfo{person}{Manuela Ver{\'\i}ssimo}.} \bibinfo{year}{2021}\natexlab{}.
\newblock \showarticletitle{Father involvement during early childhood: A systematic review of the literature}.
\newblock \bibinfo{journal}{\emph{Journal of Family Theory \& Review}} \bibinfo{volume}{13}, \bibinfo{number}{1} (\bibinfo{year}{2021}), \bibinfo{pages}{77--99}.
\newblock


\bibitem[Dourish and Bellotti(1992)]%
        {dourish1992awareness}
\bibfield{author}{\bibinfo{person}{Paul Dourish} {and} \bibinfo{person}{Victoria Bellotti}.} \bibinfo{year}{1992}\natexlab{}.
\newblock \showarticletitle{Awareness and coordination in shared workspaces}. In \bibinfo{booktitle}{\emph{Proceedings of the 1992 ACM conference on Computer-supported cooperative work}}. \bibinfo{pages}{107--114}.
\newblock


\bibitem[Dworkin et~al\mbox{.}(2016)]%
        {dworkin2016coparenting}
\bibfield{author}{\bibinfo{person}{Jodi Dworkin}, \bibinfo{person}{Ellie McCann}, {and} \bibinfo{person}{Jenifer~K McGuire}.} \bibinfo{year}{2016}\natexlab{}.
\newblock \showarticletitle{Coparenting in the digital era: Exploring divorced parents’ use of technology}.
\newblock In \bibinfo{booktitle}{\emph{Divorce, separation, and remarriage: The transformation of family}}. \bibinfo{publisher}{Emerald Group Publishing Limited}.
\newblock


\bibitem[Eichhorn et~al\mbox{.}(2008)]%
        {eichhorn2008stroking}
\bibfield{author}{\bibinfo{person}{Elisabeth Eichhorn}, \bibinfo{person}{Reto Wettach}, {and} \bibinfo{person}{Eva Hornecker}.} \bibinfo{year}{2008}\natexlab{}.
\newblock \showarticletitle{A stroking device for spatially separated couples}. In \bibinfo{booktitle}{\emph{Proceedings of the 10th international conference on Human computer interaction with mobile devices and services}}. \bibinfo{pages}{303--306}.
\newblock


\bibitem[Eira~Nunes et~al\mbox{.}(2021)]%
        {eira2021co}
\bibfield{author}{\bibinfo{person}{Cindy Eira~Nunes}, \bibinfo{person}{Yves De~Roten}, \bibinfo{person}{Nahema El~Ghaziri}, \bibinfo{person}{Nicolas Favez}, {and} \bibinfo{person}{Jo{\"e}lle Darwiche}.} \bibinfo{year}{2021}\natexlab{}.
\newblock \showarticletitle{Co-parenting programs: A systematic review and meta-analysis}.
\newblock \bibinfo{journal}{\emph{Family Relations}} \bibinfo{volume}{70}, \bibinfo{number}{3} (\bibinfo{year}{2021}), \bibinfo{pages}{759--776}.
\newblock


\bibitem[Ellegaard~Christensen et~al\mbox{.}(2021)]%
        {ellegaard2021shaping}
\bibfield{author}{\bibinfo{person}{Anne Ellegaard~Christensen}, \bibinfo{person}{Malene~H Magnussen}, \bibinfo{person}{Tobias~S Seindal}, {and} \bibinfo{person}{Dimitrios Raptis}.} \bibinfo{year}{2021}\natexlab{}.
\newblock \showarticletitle{Shaping Romance: Mediating Intimacy for Co-Located Couples}. In \bibinfo{booktitle}{\emph{Proceedings of the 33rd Australian Conference on Human-Computer Interaction}}. \bibinfo{pages}{86--98}.
\newblock


\bibitem[Entsieh and Hallstr{\"o}m(2016)]%
        {entsieh2016first}
\bibfield{author}{\bibinfo{person}{Angela~Afua Entsieh} {and} \bibinfo{person}{Inger~Kristensson Hallstr{\"o}m}.} \bibinfo{year}{2016}\natexlab{}.
\newblock \showarticletitle{First-time parents’ prenatal needs for early parenthood preparation-A systematic review and meta-synthesis of qualitative literature}.
\newblock \bibinfo{journal}{\emph{Midwifery}}  \bibinfo{volume}{39} (\bibinfo{year}{2016}), \bibinfo{pages}{1--11}.
\newblock


\bibitem[Feinberg(2002)]%
        {feinberg2002coparenting}
\bibfield{author}{\bibinfo{person}{Mark~E Feinberg}.} \bibinfo{year}{2002}\natexlab{}.
\newblock \showarticletitle{Coparenting and the transition to parenthood: A framework for prevention}.
\newblock \bibinfo{journal}{\emph{Clinical child and family psychology review}} \bibinfo{volume}{5}, \bibinfo{number}{3} (\bibinfo{year}{2002}), \bibinfo{pages}{173--195}.
\newblock


\bibitem[Feinberg(2003)]%
        {feinberg2003internal}
\bibfield{author}{\bibinfo{person}{Mark~E Feinberg}.} \bibinfo{year}{2003}\natexlab{}.
\newblock \showarticletitle{The internal structure and ecological context of coparenting: A framework for research and intervention}.
\newblock \bibinfo{journal}{\emph{Parenting: Science and Practice}} \bibinfo{volume}{3}, \bibinfo{number}{2} (\bibinfo{year}{2003}), \bibinfo{pages}{95--131}.
\newblock


\bibitem[Foundation(2020)]%
        {childwelfarefoundation2020father}
\bibfield{author}{\bibinfo{person}{Child Welfare~League Foundation}.} \bibinfo{year}{2020}\natexlab{}.
\newblock \bibinfo{booktitle}{\emph{2020 Survey Report on Taiwanese Men’s Parenting Attitudes and Current Situation}}.
\newblock
\urldef\tempurl%
\url{https://www.children.org.tw/publication_research/research_report/340}
\showURL{%
\tempurl}


\bibitem[Futris et~al\mbox{.}(2013)]%
        {futris2013national}
\bibfield{author}{\bibinfo{person}{TG Futris}, \bibinfo{person}{F Adler-Baeder}, {et~al\mbox{.}}} \bibinfo{year}{2013}\natexlab{}.
\newblock \showarticletitle{The National Extension Relationship and Marriage Education Model: Core teaching concepts for relationship and marriage enrichment programming}.
\newblock \bibinfo{journal}{\emph{Athens, GA: The University of Georgia Cooperative Extension. Retrieved February}}  \bibinfo{volume}{21} (\bibinfo{year}{2013}), \bibinfo{pages}{2016}.
\newblock


\bibitem[Gordon et~al\mbox{.}(2012)]%
        {gordon2012have}
\bibfield{author}{\bibinfo{person}{Amie~M Gordon}, \bibinfo{person}{Emily~A Impett}, \bibinfo{person}{Aleksandr Kogan}, \bibinfo{person}{Christopher Oveis}, {and} \bibinfo{person}{Dacher Keltner}.} \bibinfo{year}{2012}\natexlab{}.
\newblock \showarticletitle{To have and to hold: gratitude promotes relationship maintenance in intimate bonds.}
\newblock \bibinfo{journal}{\emph{Journal of personality and social psychology}} \bibinfo{volume}{103}, \bibinfo{number}{2} (\bibinfo{year}{2012}), \bibinfo{pages}{257}.
\newblock


\bibitem[Gordon et~al\mbox{.}(2011)]%
        {gordon2011have}
\bibfield{author}{\bibinfo{person}{Cameron~L Gordon}, \bibinfo{person}{Robyn~AM Arnette}, {and} \bibinfo{person}{Rachel~E Smith}.} \bibinfo{year}{2011}\natexlab{}.
\newblock \showarticletitle{Have you thanked your spouse today?: Felt and expressed gratitude among married couples}.
\newblock \bibinfo{journal}{\emph{Personality and Individual Differences}} \bibinfo{volume}{50}, \bibinfo{number}{3} (\bibinfo{year}{2011}), \bibinfo{pages}{339--343}.
\newblock


\bibitem[Gottman(1998)]%
        {gottman1998psychology}
\bibfield{author}{\bibinfo{person}{John~Mordechai Gottman}.} \bibinfo{year}{1998}\natexlab{}.
\newblock \showarticletitle{Psychology and the study of marital processes}.
\newblock \bibinfo{journal}{\emph{Annual review of psychology}} \bibinfo{volume}{49}, \bibinfo{number}{1} (\bibinfo{year}{1998}), \bibinfo{pages}{169--197}.
\newblock


\bibitem[Gottman and Notarius(2000)]%
        {gottman2000decade}
\bibfield{author}{\bibinfo{person}{John~M Gottman} {and} \bibinfo{person}{Clifford~I Notarius}.} \bibinfo{year}{2000}\natexlab{}.
\newblock \showarticletitle{Decade review: Observing marital interaction}.
\newblock \bibinfo{journal}{\emph{Journal of marriage and family}} \bibinfo{volume}{62}, \bibinfo{number}{4} (\bibinfo{year}{2000}), \bibinfo{pages}{927--947}.
\newblock


\bibitem[Grant et~al\mbox{.}(2016)]%
        {grant2016intervention}
\bibfield{author}{\bibinfo{person}{N Grant}, \bibinfo{person}{S Rodger}, {and} \bibinfo{person}{T Hoffmann}.} \bibinfo{year}{2016}\natexlab{}.
\newblock \showarticletitle{Intervention decision-making processes and information preferences of parents of children with autism spectrum disorders}.
\newblock \bibinfo{journal}{\emph{Child: care, health and development}} \bibinfo{volume}{42}, \bibinfo{number}{1} (\bibinfo{year}{2016}), \bibinfo{pages}{125--134}.
\newblock


\bibitem[Hazan and Shaver(1994)]%
        {hazan1994attachment}
\bibfield{author}{\bibinfo{person}{Cindy Hazan} {and} \bibinfo{person}{Phillip~R Shaver}.} \bibinfo{year}{1994}\natexlab{}.
\newblock \showarticletitle{Attachment as an organizational framework for research on close relationships}.
\newblock \bibinfo{journal}{\emph{Psychological inquiry}} \bibinfo{volume}{5}, \bibinfo{number}{1} (\bibinfo{year}{1994}), \bibinfo{pages}{1--22}.
\newblock


\bibitem[He(2013)]%
        {he2013couple}
\bibfield{author}{\bibinfo{person}{Li He}.} \bibinfo{year}{2013}\natexlab{}.
\newblock \showarticletitle{Couple collaboration: a design research exploration}.
\newblock In \bibinfo{booktitle}{\emph{CHI'13 Extended Abstracts on Human Factors in Computing Systems}}. \bibinfo{pages}{2689--2694}.
\newblock


\bibitem[Ho et~al\mbox{.}(2010)]%
        {ho2010parental}
\bibfield{author}{\bibinfo{person}{Hsiu-Zu Ho}, \bibinfo{person}{Wei-Wen Chen}, \bibinfo{person}{Connie~N Tran}, {and} \bibinfo{person}{Chu-Ting Ko}.} \bibinfo{year}{2010}\natexlab{}.
\newblock \showarticletitle{Parental involvement in Taiwanese families: Father-mother differences}.
\newblock \bibinfo{journal}{\emph{Childhood Education}} \bibinfo{volume}{86}, \bibinfo{number}{6} (\bibinfo{year}{2010}), \bibinfo{pages}{376--381}.
\newblock


\bibitem[Ho and Lam(2019)]%
        {ho2019father}
\bibfield{author}{\bibinfo{person}{Hsiu-Zu Ho} {and} \bibinfo{person}{Yeana~W Lam}.} \bibinfo{year}{2019}\natexlab{}.
\newblock \showarticletitle{Father involvement in East Asia: Beyond the breadwinner role?}
\newblock \bibinfo{journal}{\emph{The Wiley handbook of family, school, and community relationships in education}} (\bibinfo{year}{2019}), \bibinfo{pages}{333--356}.
\newblock


\bibitem[Hsiu-Zu et~al\mbox{.}(2011)]%
        {hsiu2011parent}
\bibfield{author}{\bibinfo{person}{Ho Hsiu-Zu}, \bibinfo{person}{Connie~N Tran}, \bibinfo{person}{Chu-Ting Ko}, \bibinfo{person}{Jessica~M Phillips}, \bibinfo{person}{Alma Boutin-Martinez}, \bibinfo{person}{Carol~N Dixon}, {and} \bibinfo{person}{Wei-Wen Chen}.} \bibinfo{year}{2011}\natexlab{}.
\newblock \showarticletitle{Parent involvement: Voices of Taiwanese fathers}.
\newblock \bibinfo{journal}{\emph{International Journal about Parents in Education}} \bibinfo{volume}{5}, \bibinfo{number}{2} (\bibinfo{year}{2011}).
\newblock


\bibitem[Huang et~al\mbox{.}(2022)]%
        {huang2022maternal}
\bibfield{author}{\bibinfo{person}{Hsin-Hui Huang}, \bibinfo{person}{Tzu-Ying Lee}, \bibinfo{person}{Xin-Ting Lin}, {and} \bibinfo{person}{Hui-Ying Duan}.} \bibinfo{year}{2022}\natexlab{}.
\newblock \showarticletitle{Maternal confidence and parenting stress of first-time mothers in Taiwan: The impact of sources and types of social support}. In \bibinfo{booktitle}{\emph{Healthcare}}, Vol.~\bibinfo{volume}{10}. MDPI, \bibinfo{pages}{878}.
\newblock


\bibitem[Jensen et~al\mbox{.}(2019)]%
        {jensen2019into}
\bibfield{author}{\bibinfo{person}{Jonas~Kjeldmand Jensen}, \bibinfo{person}{Tawfiq Ammari}, {and} \bibinfo{person}{Pernille Bj{\o}rn}.} \bibinfo{year}{2019}\natexlab{}.
\newblock \showarticletitle{Into Scandinavia: When Online Fatherhood Reflects Societal Infrastructures}.
\newblock \bibinfo{journal}{\emph{Proceedings of the ACM on human-computer interaction}} \bibinfo{volume}{3}, \bibinfo{number}{GROUP} (\bibinfo{year}{2019}), \bibinfo{pages}{1--21}.
\newblock


\bibitem[Jo et~al\mbox{.}(2020)]%
        {jo2020understanding}
\bibfield{author}{\bibinfo{person}{Eunkyung Jo}, \bibinfo{person}{Austin~L Toombs}, \bibinfo{person}{Colin~M Gray}, {and} \bibinfo{person}{Hwajung Hong}.} \bibinfo{year}{2020}\natexlab{}.
\newblock \showarticletitle{Understanding parenting stress through co-designed self-trackers}. In \bibinfo{booktitle}{\emph{Proceedings of the 2020 CHI Conference on Human Factors in Computing Systems}}. \bibinfo{pages}{1--13}.
\newblock


\bibitem[Karandashev and Karandashev(2019)]%
        {karandashev2019love}
\bibfield{author}{\bibinfo{person}{Victor Karandashev} {and} \bibinfo{person}{Victor Karandashev}.} \bibinfo{year}{2019}\natexlab{}.
\newblock \showarticletitle{Love as Intimacy}.
\newblock \bibinfo{journal}{\emph{Cross-Cultural Perspectives on the Experience and Expression of Love}} (\bibinfo{year}{2019}), \bibinfo{pages}{187--202}.
\newblock


\bibitem[Khazan et~al\mbox{.}(2008)]%
        {khazan2008violated}
\bibfield{author}{\bibinfo{person}{Inna Khazan}, \bibinfo{person}{James~P Mchale}, {and} \bibinfo{person}{Wendy Decourcey}.} \bibinfo{year}{2008}\natexlab{}.
\newblock \showarticletitle{Violated wishes about division of childcare labor predict early coparenting process during stressful and nonstressful family evaluations}.
\newblock \bibinfo{journal}{\emph{Infant Mental Health Journal: Official Publication of The World Association for Infant Mental Health}} \bibinfo{volume}{29}, \bibinfo{number}{4} (\bibinfo{year}{2008}), \bibinfo{pages}{343--361}.
\newblock


\bibitem[Kirchner et~al\mbox{.}(2020)]%
        {kirchner2020just}
\bibfield{author}{\bibinfo{person}{Susanne Kirchner}, \bibinfo{person}{Dawn~K Sakaguchi-Tang}, \bibinfo{person}{Rebecca Michelson}, \bibinfo{person}{Sean~A Munson}, {and} \bibinfo{person}{Julie~A Kientz}.} \bibinfo{year}{2020}\natexlab{}.
\newblock \showarticletitle{"This just felt to me like the right thing to do" Decision-Making Experiences of Parents of Young Children}. In \bibinfo{booktitle}{\emph{Proceedings of the 2020 ACM designing interactive systems conference}}. \bibinfo{pages}{489--503}.
\newblock


\bibitem[Ko and Ma(2022)]%
        {ko2022mobilizing}
\bibfield{author}{\bibinfo{person}{Man~Ching Ko} {and} \bibinfo{person}{Xiaojuan Ma}.} \bibinfo{year}{2022}\natexlab{}.
\newblock \showarticletitle{Mobilizing Instrumental Childcare Support for Postpartum Mothers: Needs for and Barriers to Infant-centric Family Informatics Practices in Hong Kong}.
\newblock \bibinfo{journal}{\emph{Proceedings of the ACM on Human-Computer Interaction}} \bibinfo{volume}{6}, \bibinfo{number}{CSCW2} (\bibinfo{year}{2022}), \bibinfo{pages}{1--40}.
\newblock


\bibitem[Kowalski et~al\mbox{.}(2013)]%
        {kowalski2013cubble}
\bibfield{author}{\bibinfo{person}{Robert Kowalski}, \bibinfo{person}{Sebastian Loehmann}, {and} \bibinfo{person}{Doris Hausen}.} \bibinfo{year}{2013}\natexlab{}.
\newblock \showarticletitle{Cubble: A multi-device hybrid approach supporting communication in long-distance relationships}. In \bibinfo{booktitle}{\emph{Proceedings of the 7th International Conference on Tangible, Embedded and Embodied Interaction}}. \bibinfo{pages}{201--204}.
\newblock


\bibitem[Kwon et~al\mbox{.}(2013)]%
        {kwon2013mothers}
\bibfield{author}{\bibinfo{person}{Kyong-Ah Kwon}, \bibinfo{person}{Suejung Han}, \bibinfo{person}{Hyun-Joo Jeon}, {and} \bibinfo{person}{Gary~E Bingham}.} \bibinfo{year}{2013}\natexlab{}.
\newblock \showarticletitle{Mothers' and fathers' parenting challenges, strategies, and resources in toddlerhood}.
\newblock \bibinfo{journal}{\emph{Early Child Development and Care}} \bibinfo{volume}{183}, \bibinfo{number}{3-4} (\bibinfo{year}{2013}), \bibinfo{pages}{415--429}.
\newblock


\bibitem[Laurenceau et~al\mbox{.}(2005)]%
        {laurenceau2005interpersonal}
\bibfield{author}{\bibinfo{person}{Jean-Philippe Laurenceau}, \bibinfo{person}{Lisa~Feldman Barrett}, {and} \bibinfo{person}{Michael~J Rovine}.} \bibinfo{year}{2005}\natexlab{}.
\newblock \showarticletitle{The interpersonal process model of intimacy in marriage: a daily-diary and multilevel modeling approach.}
\newblock \bibinfo{journal}{\emph{Journal of family psychology}} \bibinfo{volume}{19}, \bibinfo{number}{2} (\bibinfo{year}{2005}), \bibinfo{pages}{314}.
\newblock


\bibitem[Le et~al\mbox{.}(2019)]%
        {le2019cross}
\bibfield{author}{\bibinfo{person}{Yunying Le}, \bibinfo{person}{Steffany~J Fredman}, \bibinfo{person}{Brandon~T McDaniel}, \bibinfo{person}{Jean-Philippe Laurenceau}, {and} \bibinfo{person}{Mark~E Feinberg}.} \bibinfo{year}{2019}\natexlab{}.
\newblock \showarticletitle{Cross-day influences between couple closeness and coparenting support among new parents.}
\newblock \bibinfo{journal}{\emph{Journal of Family Psychology}} \bibinfo{volume}{33}, \bibinfo{number}{3} (\bibinfo{year}{2019}), \bibinfo{pages}{360}.
\newblock


\bibitem[LeMasters(1957)]%
        {lemasters1957parenthood}
\bibfield{author}{\bibinfo{person}{Ersel~E LeMasters}.} \bibinfo{year}{1957}\natexlab{}.
\newblock \showarticletitle{Parenthood as crisis}.
\newblock \bibinfo{journal}{\emph{Marriage and family living}} \bibinfo{volume}{19}, \bibinfo{number}{4} (\bibinfo{year}{1957}), \bibinfo{pages}{352--355}.
\newblock


\bibitem[Lin et~al\mbox{.}(2024)]%
        {lin2024ultimately}
\bibfield{author}{\bibinfo{person}{Ya-Fang Lin}, \bibinfo{person}{Na Li}, \bibinfo{person}{Wan-Hsuan Huang}, \bibinfo{person}{Karen Ecsedy}, \bibinfo{person}{Mark~E Feinberg}, \bibinfo{person}{Douglas Teti}, {and} \bibinfo{person}{John~M Carroll}.} \bibinfo{year}{2024}\natexlab{}.
\newblock \showarticletitle{" Ultimately We're Together": Understanding New Parents' Experiences of Co-parenting}.
\newblock \bibinfo{journal}{\emph{Proceedings of the ACM on Human-Computer Interaction}} \bibinfo{volume}{8}, \bibinfo{number}{CSCW2} (\bibinfo{year}{2024}), \bibinfo{pages}{1--25}.
\newblock


\bibitem[Lindquist et~al\mbox{.}(2007)]%
        {lindquist2007co}
\bibfield{author}{\bibinfo{person}{Sinna Lindquist}, \bibinfo{person}{Bo Westerlund}, \bibinfo{person}{Yngve Sundblad}, \bibinfo{person}{Helena Tobiasson}, \bibinfo{person}{Michel Beaudouin-Lafon}, {and} \bibinfo{person}{Wendy Mackay}.} \bibinfo{year}{2007}\natexlab{}.
\newblock \showarticletitle{Co-designing communication technology with and for families--Methods, experience, results and impact}.
\newblock \bibinfo{journal}{\emph{The Disappearing Computer: Interaction Design, System Infrastructures and Applications for Smart Environments}} (\bibinfo{year}{2007}), \bibinfo{pages}{99--119}.
\newblock


\bibitem[Lu et~al\mbox{.}(2024a)]%
        {lu2024examining}
\bibfield{author}{\bibinfo{person}{Xi Lu}, \bibinfo{person}{Jacquelyn~E Powell}, \bibinfo{person}{Elena Agapie}, \bibinfo{person}{Yunan Chen}, {and} \bibinfo{person}{Daniel~A Epstein}.} \bibinfo{year}{2024}\natexlab{a}.
\newblock \showarticletitle{Examining the Social Aspects of Pregnancy Tracking Applications}.
\newblock \bibinfo{journal}{\emph{Proceedings of the ACM on Human-Computer Interaction}} \bibinfo{volume}{8}, \bibinfo{number}{CSCW1} (\bibinfo{year}{2024}), \bibinfo{pages}{1--30}.
\newblock


\bibitem[Lu et~al\mbox{.}(2024b)]%
        {lu2024unpacking}
\bibfield{author}{\bibinfo{person}{Xi Lu}, \bibinfo{person}{Jacquelyn~E Powell}, \bibinfo{person}{Elena Agapie}, \bibinfo{person}{Yunan Chen}, {and} \bibinfo{person}{Daniel~A Epstein}.} \bibinfo{year}{2024}\natexlab{b}.
\newblock \showarticletitle{Unpacking the Lived Experience of Collaborative Pregnancy Tracking}. In \bibinfo{booktitle}{\emph{Proceedings of the CHI Conference on Human Factors in Computing Systems}}. \bibinfo{pages}{1--17}.
\newblock


\bibitem[Lucier-Greer et~al\mbox{.}(2018)]%
        {lucier2018enhancing}
\bibfield{author}{\bibinfo{person}{Mallory Lucier-Greer}, \bibinfo{person}{Amelia~J Birney}, \bibinfo{person}{Teri~M Gutierrez}, {and} \bibinfo{person}{Francesca Adler-Baeder}.} \bibinfo{year}{2018}\natexlab{}.
\newblock \showarticletitle{Enhancing relationship skills and couple functioning with mobile technology: An evaluation of the Love Every Day mobile intervention}.
\newblock \bibinfo{journal}{\emph{Journal of Family Social Work}} \bibinfo{volume}{21}, \bibinfo{number}{2} (\bibinfo{year}{2018}), \bibinfo{pages}{152--171}.
\newblock


\bibitem[Lukoff et~al\mbox{.}(2017)]%
        {lukoff2017gender}
\bibfield{author}{\bibinfo{person}{Kai Lukoff}, \bibinfo{person}{Carol Moser}, {and} \bibinfo{person}{Sarita Schoenebeck}.} \bibinfo{year}{2017}\natexlab{}.
\newblock \showarticletitle{Gender norms and attitudes about childcare activities presented on father blogs}. In \bibinfo{booktitle}{\emph{Proceedings of the 2017 CHI Conference on Human Factors in Computing Systems}}. \bibinfo{pages}{4966--4971}.
\newblock


\bibitem[McHale et~al\mbox{.}(2004)]%
        {mchale2004growing}
\bibfield{author}{\bibinfo{person}{James~P McHale}, \bibinfo{person}{Regina Kuersten-Hogan}, {and} \bibinfo{person}{Nirmala Rao}.} \bibinfo{year}{2004}\natexlab{}.
\newblock \showarticletitle{Growing points for coparenting theory and research}.
\newblock \bibinfo{journal}{\emph{Journal of adult development}} \bibinfo{volume}{11}, \bibinfo{number}{3} (\bibinfo{year}{2004}), \bibinfo{pages}{221--234}.
\newblock


\bibitem[Newland et~al\mbox{.}(2013)]%
        {newland2013parent}
\bibfield{author}{\bibinfo{person}{Lisa~A Newland}, \bibinfo{person}{Hui-Hua Chen}, \bibinfo{person}{Diana~D Coyl-Shepherd}, \bibinfo{person}{Yi-Ching Liang}, \bibinfo{person}{Eliann~R Carr}, \bibinfo{person}{Emily Dykstra}, {and} \bibinfo{person}{Susan~C Gapp}.} \bibinfo{year}{2013}\natexlab{}.
\newblock \showarticletitle{Parent and child perspectives on mothering and fathering: The influence of ecocultural niches}.
\newblock \bibinfo{journal}{\emph{Early Child Development and Care}} \bibinfo{volume}{183}, \bibinfo{number}{3-4} (\bibinfo{year}{2013}), \bibinfo{pages}{534--552}.
\newblock


\bibitem[Nikkhah et~al\mbox{.}(2022)]%
        {nikkhah2022family}
\bibfield{author}{\bibinfo{person}{Sarah Nikkhah}, \bibinfo{person}{Swaroop John}, \bibinfo{person}{Krishna~Supradeep Yalamarti}, \bibinfo{person}{Emily~L Mueller}, {and} \bibinfo{person}{Andrew~D Miller}.} \bibinfo{year}{2022}\natexlab{}.
\newblock \showarticletitle{Family Care Coordination in the Children's Hospital: Phases and Cycles in the Pediatric Cancer Caregiving Journey}.
\newblock \bibinfo{journal}{\emph{Proceedings of the ACM on Human-Computer Interaction}} \bibinfo{volume}{6}, \bibinfo{number}{CSCW2} (\bibinfo{year}{2022}), \bibinfo{pages}{1--30}.
\newblock


\bibitem[Odom et~al\mbox{.}(2010)]%
        {odom2010designing}
\bibfield{author}{\bibinfo{person}{William Odom}, \bibinfo{person}{John Zimmerman}, {and} \bibinfo{person}{Jodi Forlizzi}.} \bibinfo{year}{2010}\natexlab{}.
\newblock \showarticletitle{Designing for dynamic family structures: divorced families and interactive systems}. In \bibinfo{booktitle}{\emph{Proceedings of the 8th ACM conference on designing interactive systems}}. \bibinfo{pages}{151--160}.
\newblock


\bibitem[Park et~al\mbox{.}(2023)]%
        {park2023actor}
\bibfield{author}{\bibinfo{person}{In~Young Park}, \bibinfo{person}{Jennifer~L Bellamy}, \bibinfo{person}{Stephanie~Rachel Speer}, \bibinfo{person}{Jangmin Kim}, \bibinfo{person}{Jin~Yao Kwan}, \bibinfo{person}{Paula Powe}, \bibinfo{person}{Aaron Banman}, \bibinfo{person}{Justin~S Harty}, {and} \bibinfo{person}{Neil~B Guterman}.} \bibinfo{year}{2023}\natexlab{}.
\newblock \showarticletitle{Actor and Partner Effects of Interparental Relationship and Co-Parenting on Parenting Stress Among Fathers and Mothers}.
\newblock \bibinfo{journal}{\emph{Families in Society}} (\bibinfo{year}{2023}), \bibinfo{pages}{10443894231207442}.
\newblock


\bibitem[Pfeifer et~al\mbox{.}(2013)]%
        {pfeifer2013perceived}
\bibfield{author}{\bibinfo{person}{Lexie Pfeifer}, \bibinfo{person}{Richard~B Miller}, \bibinfo{person}{Tsui-Shan Li}, {and} \bibinfo{person}{Ying-Ling Hsiao}.} \bibinfo{year}{2013}\natexlab{}.
\newblock \showarticletitle{Perceived marital problems in Taiwan}.
\newblock \bibinfo{journal}{\emph{Contemporary Family Therapy}}  \bibinfo{volume}{35} (\bibinfo{year}{2013}), \bibinfo{pages}{91--104}.
\newblock


\bibitem[Pietromonaco and Collins(2017)]%
        {pietromonaco2017interpersonal}
\bibfield{author}{\bibinfo{person}{Paula~R Pietromonaco} {and} \bibinfo{person}{Nancy~L Collins}.} \bibinfo{year}{2017}\natexlab{}.
\newblock \showarticletitle{Interpersonal mechanisms linking close relationships to health.}
\newblock \bibinfo{journal}{\emph{American Psychologist}} \bibinfo{volume}{72}, \bibinfo{number}{6} (\bibinfo{year}{2017}), \bibinfo{pages}{531}.
\newblock


\bibitem[Ramchandani et~al\mbox{.}(2013)]%
        {ramchandani2013early}
\bibfield{author}{\bibinfo{person}{Paul~G Ramchandani}, \bibinfo{person}{Jill Domoney}, \bibinfo{person}{Vaheshta Sethna}, \bibinfo{person}{Lamprini Psychogiou}, \bibinfo{person}{Haido Vlachos}, {and} \bibinfo{person}{Lynne Murray}.} \bibinfo{year}{2013}\natexlab{}.
\newblock \showarticletitle{Do early father--infant interactions predict the onset of externalising behaviours in young children? Findings from a longitudinal cohort study}.
\newblock \bibinfo{journal}{\emph{Journal of child psychology and psychiatry}} \bibinfo{volume}{54}, \bibinfo{number}{1} (\bibinfo{year}{2013}), \bibinfo{pages}{56--64}.
\newblock


\bibitem[Reis et~al\mbox{.}(2018)]%
        {reis2018intimacy}
\bibfield{author}{\bibinfo{person}{Harry~T Reis} {et~al\mbox{.}}} \bibinfo{year}{2018}\natexlab{}.
\newblock \showarticletitle{Intimacy as an interpersonal process}.
\newblock In \bibinfo{booktitle}{\emph{Relationships, well-being and behaviour}}. \bibinfo{publisher}{Routledge}, \bibinfo{pages}{113--143}.
\newblock


\bibitem[Riggs and Bartholomaeus(2020)]%
        {riggs2020s}
\bibfield{author}{\bibinfo{person}{Damien~W Riggs} {and} \bibinfo{person}{Clare Bartholomaeus}.} \bibinfo{year}{2020}\natexlab{}.
\newblock \showarticletitle{‘That’s my job’: Accounting for division of labour amongst heterosexual first time parents}.
\newblock \bibinfo{journal}{\emph{Community, Work \& Family}} \bibinfo{volume}{23}, \bibinfo{number}{1} (\bibinfo{year}{2020}), \bibinfo{pages}{107--122}.
\newblock


\bibitem[Rosenfeld and Hausen(2023)]%
        {rosenfeld2023resilience}
\bibfield{author}{\bibinfo{person}{Michael~J Rosenfeld} {and} \bibinfo{person}{Sonia Hausen}.} \bibinfo{year}{2023}\natexlab{}.
\newblock \showarticletitle{Resilience and Stress in Romantic Relationships in the United States During the COVID-19 Pandemic}.
\newblock \bibinfo{journal}{\emph{Sociological Science}}  \bibinfo{volume}{10} (\bibinfo{year}{2023}), \bibinfo{pages}{472--500}.
\newblock


\bibitem[Rusbult et~al\mbox{.}(2004)]%
        {rusbult2004interdependence}
\bibfield{author}{\bibinfo{person}{Caryl~E Rusbult}, \bibinfo{person}{Madoka Kumashiro}, \bibinfo{person}{Michael~K Coolsen}, {and} \bibinfo{person}{Jeffrey~L Kirchner}.} \bibinfo{year}{2004}\natexlab{}.
\newblock \showarticletitle{Interdependence, closeness, and relationships}.
\newblock \bibinfo{journal}{\emph{Handbook of closeness and intimacy}} (\bibinfo{year}{2004}), \bibinfo{pages}{147--172}.
\newblock


\bibitem[Sanders and Stappers(2008)]%
        {sanders2008co}
\bibfield{author}{\bibinfo{person}{Elizabeth B-N Sanders} {and} \bibinfo{person}{Pieter~Jan Stappers}.} \bibinfo{year}{2008}\natexlab{}.
\newblock \showarticletitle{Co-creation and the new landscapes of design}.
\newblock \bibinfo{journal}{\emph{Co-design}} \bibinfo{volume}{4}, \bibinfo{number}{1} (\bibinfo{year}{2008}), \bibinfo{pages}{5--18}.
\newblock


\bibitem[Scherer(2021)]%
        {USCensus2021}
\bibfield{author}{\bibinfo{person}{Zachary Scherer}.} \bibinfo{year}{September 16, 2021}\natexlab{}.
\newblock \bibinfo{title}{College-Educated Women and Non-Hispanic White Women More Likely to Work During First Pregnancy}.
\newblock
\newblock
\urldef\tempurl%
\url{https://www.census.gov/library/stories/2021/09/two-thirds-recent-first-time-fathers-took-time-off-after-birth.html#:~:text=Roughly%20two-thirds%20of%20men%20who%20were%20fathers%20for,according%20to%20U.S.%20Census%20Bureau%20data%20released%20today.}
\showURL{%
\tempurl}


\bibitem[Setiawan et~al\mbox{.}(2022)]%
        {setiawan2022understanding}
\bibfield{author}{\bibinfo{person}{Jenny~Lukito Setiawan}, \bibinfo{person}{Jessica~Christina Widhigdo}, \bibinfo{person}{Lisa Indriati}, \bibinfo{person}{Mychael~Maoeretz Engel}, {and} \bibinfo{person}{Amanda Teonata}.} \bibinfo{year}{2022}\natexlab{}.
\newblock \showarticletitle{Understanding the Issues of Co-parenting in Indonesia}.
\newblock  (\bibinfo{year}{2022}).
\newblock


\bibitem[Shapiro et~al\mbox{.}(2000)]%
        {shapiro2000baby}
\bibfield{author}{\bibinfo{person}{Alyson~Fearnley Shapiro}, \bibinfo{person}{John~M Gottman}, {and} \bibinfo{person}{Sybil Carrere}.} \bibinfo{year}{2000}\natexlab{}.
\newblock \showarticletitle{The baby and the marriage: identifying factors that buffer against decline in marital satisfaction after the first baby arrives.}
\newblock \bibinfo{journal}{\emph{Journal of family psychology}} \bibinfo{volume}{14}, \bibinfo{number}{1} (\bibinfo{year}{2000}), \bibinfo{pages}{59}.
\newblock


\bibitem[Sheedy and Gambrel(2019)]%
        {sheedy2019coparenting}
\bibfield{author}{\bibinfo{person}{Anna Sheedy} {and} \bibinfo{person}{Laura~Eubanks Gambrel}.} \bibinfo{year}{2019}\natexlab{}.
\newblock \showarticletitle{Coparenting negotiation during the transition to parenthood: A qualitative study of couples’ experiences as new parents}.
\newblock \bibinfo{journal}{\emph{The American Journal of Family Therapy}} \bibinfo{volume}{47}, \bibinfo{number}{2} (\bibinfo{year}{2019}), \bibinfo{pages}{67--86}.
\newblock


\bibitem[Shen(2005)]%
        {shen2005factors}
\bibfield{author}{\bibinfo{person}{April Chiung-Tao Shen}.} \bibinfo{year}{2005}\natexlab{}.
\newblock \showarticletitle{Factors in the marital relationship in a changing society: A Taiwan case study}.
\newblock \bibinfo{journal}{\emph{International Social Work}} \bibinfo{volume}{48}, \bibinfo{number}{3} (\bibinfo{year}{2005}), \bibinfo{pages}{325--340}.
\newblock


\bibitem[Shin et~al\mbox{.}(2022)]%
        {shin2022more}
\bibfield{author}{\bibinfo{person}{Ji~Youn Shin}, \bibinfo{person}{Wei Peng}, {and} \bibinfo{person}{Hee~Rin Lee}.} \bibinfo{year}{2022}\natexlab{}.
\newblock \showarticletitle{More than Bedtime and the Bedroom: Sleep Management as a Collaborative Work for the Family}. In \bibinfo{booktitle}{\emph{CHI Conference on Human Factors in Computing Systems}}. \bibinfo{pages}{1--16}.
\newblock


\bibitem[Smith et~al\mbox{.}(2018)]%
        {smith2018designing}
\bibfield{author}{\bibinfo{person}{Madeline~E Smith}, \bibinfo{person}{Leo Ascenzi}, \bibinfo{person}{Yingsi Qin}, {and} \bibinfo{person}{Ryan Wetsman}.} \bibinfo{year}{2018}\natexlab{}.
\newblock \showarticletitle{Designing a Video Co-Watching Web App to Support Interpersonal Relationship Maintenance}. In \bibinfo{booktitle}{\emph{Proceedings of the 2018 ACM International Conference on Supporting Group Work}}. \bibinfo{pages}{162--165}.
\newblock


\bibitem[Song et~al\mbox{.}(2018)]%
        {song2018bebecode}
\bibfield{author}{\bibinfo{person}{Seokwoo Song}, \bibinfo{person}{Juho Kim}, \bibinfo{person}{Bumsoo Kang}, \bibinfo{person}{Wonjeong Park}, {and} \bibinfo{person}{John Kim}.} \bibinfo{year}{2018}\natexlab{}.
\newblock \showarticletitle{Bebecode: Collaborative child development tracking system}. In \bibinfo{booktitle}{\emph{Proceedings of the 2018 chi conference on human factors in computing systems}}. \bibinfo{pages}{1--12}.
\newblock


\bibitem[Song et~al\mbox{.}(2020)]%
        {song2020bodeum}
\bibfield{author}{\bibinfo{person}{Seokwoo Song}, \bibinfo{person}{Naomi Yamashita}, {and} \bibinfo{person}{John Kim}.} \bibinfo{year}{2020}\natexlab{}.
\newblock \showarticletitle{Bodeum: Encouraging Working Parents to Provide Emotional Support for Stay-at-Home Parents in Korea}. In \bibinfo{booktitle}{\emph{Proceedings of the 14th EAI International Conference on Pervasive Computing Technologies for Healthcare}}. \bibinfo{pages}{38--49}.
\newblock


\bibitem[Steen(2013)]%
        {steen2013co}
\bibfield{author}{\bibinfo{person}{Marc Steen}.} \bibinfo{year}{2013}\natexlab{}.
\newblock \showarticletitle{Co-design as a process of joint inquiry and imagination}.
\newblock \bibinfo{journal}{\emph{Design issues}} \bibinfo{volume}{29}, \bibinfo{number}{2} (\bibinfo{year}{2013}), \bibinfo{pages}{16--28}.
\newblock


\bibitem[Sternberg(1986)]%
        {sternberg1986triangular}
\bibfield{author}{\bibinfo{person}{Robert~J Sternberg}.} \bibinfo{year}{1986}\natexlab{}.
\newblock \showarticletitle{A triangular theory of love.}
\newblock \bibinfo{journal}{\emph{Psychological review}} \bibinfo{volume}{93}, \bibinfo{number}{2} (\bibinfo{year}{1986}), \bibinfo{pages}{119}.
\newblock


\bibitem[Su et~al\mbox{.}(2015)]%
        {su2015cross}
\bibfield{author}{\bibinfo{person}{Li~Ping Su}, \bibinfo{person}{Richard~B Miller}, \bibinfo{person}{Jerevie~M Canlas}, \bibinfo{person}{Tsui-Shan Li}, \bibinfo{person}{Ying-Ling Hsiao}, {and} \bibinfo{person}{Brian~J Willoughby}.} \bibinfo{year}{2015}\natexlab{}.
\newblock \showarticletitle{A cross-cultural study of perceived marital problems in Taiwan and the United States}.
\newblock \bibinfo{journal}{\emph{Contemporary Family Therapy}}  \bibinfo{volume}{37} (\bibinfo{year}{2015}), \bibinfo{pages}{165--175}.
\newblock


\bibitem[Suzuki et~al\mbox{.}(2017)]%
        {suzuki2017faceshare}
\bibfield{author}{\bibinfo{person}{Keita Suzuki}, \bibinfo{person}{Masanori Yokoyama}, \bibinfo{person}{Shigeo Yoshida}, \bibinfo{person}{Takayoshi Mochizuki}, \bibinfo{person}{Tomohiro Yamada}, \bibinfo{person}{Takuji Narumi}, \bibinfo{person}{Tomohiro Tanikawa}, {and} \bibinfo{person}{Michitaka Hirose}.} \bibinfo{year}{2017}\natexlab{}.
\newblock \showarticletitle{Faceshare: Mirroring with pseudo-smile enriches video chat communications}. In \bibinfo{booktitle}{\emph{Proceedings of the 2017 CHI Conference on Human Factors in Computing Systems}}. \bibinfo{pages}{5313--5317}.
\newblock


\bibitem[Thomas et~al\mbox{.}(2018)]%
        {thomas2018appy}
\bibfield{author}{\bibinfo{person}{Gareth~M Thomas}, \bibinfo{person}{Deborah Lupton}, {and} \bibinfo{person}{Sarah Pedersen}.} \bibinfo{year}{2018}\natexlab{}.
\newblock \showarticletitle{‘The appy for a happy pappy’: expectant fatherhood and pregnancy apps}.
\newblock \bibinfo{journal}{\emph{Journal of Gender Studies}} \bibinfo{volume}{27}, \bibinfo{number}{7} (\bibinfo{year}{2018}), \bibinfo{pages}{759--770}.
\newblock


\bibitem[Tu et~al\mbox{.}(2014)]%
        {tu2014study}
\bibfield{author}{\bibinfo{person}{Yi-Chan Tu}, \bibinfo{person}{Jen-Chun Chang}, {and} \bibinfo{person}{Tsai-Feng Kao}.} \bibinfo{year}{2014}\natexlab{}.
\newblock \showarticletitle{A study on the relationships between maternal gatekeeping and paternal involvement in Taiwan}.
\newblock \bibinfo{journal}{\emph{Procedia-Social and Behavioral Sciences}}  \bibinfo{volume}{122} (\bibinfo{year}{2014}), \bibinfo{pages}{319--328}.
\newblock


\bibitem[Twenge et~al\mbox{.}(2003)]%
        {twenge2003parenthood}
\bibfield{author}{\bibinfo{person}{Jean~M Twenge}, \bibinfo{person}{W~Keith Campbell}, {and} \bibinfo{person}{Craig~A Foster}.} \bibinfo{year}{2003}\natexlab{}.
\newblock \showarticletitle{Parenthood and marital satisfaction: a meta-analytic review}.
\newblock \bibinfo{journal}{\emph{Journal of marriage and family}} \bibinfo{volume}{65}, \bibinfo{number}{3} (\bibinfo{year}{2003}), \bibinfo{pages}{574--583}.
\newblock


\bibitem[Van~Egeren and Hawkins(2004)]%
        {van2004coming}
\bibfield{author}{\bibinfo{person}{Laurie~A Van~Egeren} {and} \bibinfo{person}{Dyane~P Hawkins}.} \bibinfo{year}{2004}\natexlab{}.
\newblock \showarticletitle{Coming to terms with coparenting: Implications of definition and measurement}.
\newblock \bibinfo{journal}{\emph{Journal of Adult Development}} \bibinfo{volume}{11}, \bibinfo{number}{3} (\bibinfo{year}{2004}), \bibinfo{pages}{165--178}.
\newblock


\bibitem[van Greevenbroek et~al\mbox{.}(2023)]%
        {van2023like}
\bibfield{author}{\bibinfo{person}{Roos van Greevenbroek}, \bibinfo{person}{Dilisha Patel}, {and} \bibinfo{person}{Aneesha Singh}.} \bibinfo{year}{2023}\natexlab{}.
\newblock \showarticletitle{" Like a candy shop with forbidden fruits": Exploring Sexual Desire of Cohabiting Millennial Couples with Technology}. In \bibinfo{booktitle}{\emph{Proceedings of the 2023 ACM Designing Interactive Systems Conference}}. \bibinfo{pages}{1842--1860}.
\newblock


\bibitem[Vetere et~al\mbox{.}(2005)]%
        {vetere2005mediating}
\bibfield{author}{\bibinfo{person}{Frank Vetere}, \bibinfo{person}{Martin~R Gibbs}, \bibinfo{person}{Jesper Kjeldskov}, \bibinfo{person}{Steve Howard}, \bibinfo{person}{Florian'Floyd' Mueller}, \bibinfo{person}{Sonja Pedell}, \bibinfo{person}{Karen Mecoles}, {and} \bibinfo{person}{Marcus Bunyan}.} \bibinfo{year}{2005}\natexlab{}.
\newblock \showarticletitle{Mediating intimacy: designing technologies to support strong-tie relationships}. In \bibinfo{booktitle}{\emph{Proceedings of the SIGCHI conference on Human factors in computing systems}}. \bibinfo{pages}{471--480}.
\newblock


\bibitem[Wardle et~al\mbox{.}(2018)]%
        {wardle2018exploring}
\bibfield{author}{\bibinfo{person}{Chelsea-Joy Wardle}, \bibinfo{person}{Mitchell Green}, \bibinfo{person}{Christine~Wanjiru Mburu}, {and} \bibinfo{person}{Melissa Densmore}.} \bibinfo{year}{2018}\natexlab{}.
\newblock \showarticletitle{Exploring co-design with breastfeeding mothers}. In \bibinfo{booktitle}{\emph{Proceedings of the 2018 CHI Conference on Human Factors in Computing Systems}}. \bibinfo{pages}{1--12}.
\newblock


\bibitem[Weingarten(2022)]%
        {weingarten2022interdependence}
\bibfield{author}{\bibinfo{person}{Kathy Weingarten}.} \bibinfo{year}{2022}\natexlab{}.
\newblock \showarticletitle{Interdependence}.
\newblock In \bibinfo{booktitle}{\emph{Working couples}}. \bibinfo{publisher}{Routledge}, \bibinfo{pages}{147--158}.
\newblock


\bibitem[White et~al\mbox{.}(2024)]%
        {white2024father}
\bibfield{author}{\bibinfo{person}{Alison~C White}, \bibinfo{person}{Olivia~N Diggs}, {and} \bibinfo{person}{Tricia~K Neppl}.} \bibinfo{year}{2024}\natexlab{}.
\newblock \showarticletitle{Father and mother harsh parenting and adult romantic relationships over time: Individual behavior during adolescence.}
\newblock \bibinfo{journal}{\emph{Journal of Family Psychology}} (\bibinfo{year}{2024}).
\newblock


\bibitem[Yarosh et~al\mbox{.}(2016)]%
        {yarosh2016best}
\bibfield{author}{\bibinfo{person}{Svetlana Yarosh}, \bibinfo{person}{Sarita Schoenebeck}, \bibinfo{person}{Shreya Kothaneth}, {and} \bibinfo{person}{Elizabeth Bales}.} \bibinfo{year}{2016}\natexlab{}.
\newblock \showarticletitle{"Best of Both Worlds" Opportunities for Technology in Cross-Cultural Parenting}. In \bibinfo{booktitle}{\emph{Proceedings of the 2016 chi conference on human factors in computing systems}}. \bibinfo{pages}{635--647}.
\newblock


\end{thebibliography}

\appendix

\end{document}